\documentclass[fleqn,usenatbib]{mnras}
\usepackage{newtxtext,newtxmath}
\usepackage{float}
\usepackage{tabularx}
\usepackage{multirow}
\usepackage{longtable, lscape}
\usepackage{rotating}
\usepackage{footnote}
\usepackage{subfigure}
\usepackage{ae,aecompl}
\usepackage{graphicx}	
\usepackage{amsmath}	

\usepackage{amssymb}	
\title[EMP stars]
{Three Extremely Metal-Poor stars: discovery of a new CEMP-no star\thanks{Based [in part] on data collected at Subaru Telescope, which is operated by the National Astronomical Observatory of Japan.}\thanks{[Part of] the data are retrieved from the JVO portal (http://jvo.nao.ac.jp/portal) operated by the NAOJ}}

\author[Goswami \& Goswami]{Partha Pratim Goswami $^{1}$, Aruna Goswami $^{2}$    \\
    $^{1}$ Department of Physics, Dakshin Kamrup College, Mirza, Kamrup 781125, India;\\
    $^{2}$ Indian Institute of Astrophysics, Koramangala,  Bangalore
    560034, India;\\
    partha.goswami.iiap@gmail.com; aruna@iiap.res.in}
\date{Accepted 2026 June 1. Received 2026 June 1; in original form 2026 May 18}

\pubyear{2026}

\begin{document}
\label{firstpage}
\pagerange{\pageref{firstpage}--\pageref{lastpage}}
\maketitle

\begin{abstract}
We have conducted detailed high-resolution spectroscopic studies on three extremely metal-poor (EMP) stars HE~0401$-$0138, HE~1153$-$0518 and HE~1246$-$1344. 
For the stars HE~0401$-$0138 and HE~1246$-$1344, we have estimated the abundances of C, Na, Mg, Ca, Sc, Ti, Cr, Mn, Co, Ni, Sr and Ba along with the upper limits for Li, O, La, Ce, Pr, Nd, Sm, and Eu. For HE~1153$-$0518, abundances of seven light elements from C through Ni and two heavy elements, Y and Ba, have been derived, together with upper limits for Li, O, and La. Based on their observed abundance patterns, HE~0401$-$0138 and HE~1246$-$1344 are classified as normal EMP stars, whereas HE~1153$-$0518 is identified as a newly discovered CEMP-no star. 
HE~1153$-$0518 shows strong carbon enhancement with a high absolute carbon abundance, extreme sodium enrichment, very low neutron-capture element abundances, and a very low carbon isotopic ratio ($^{12}$C/$^{13}$C~=~2.0). Its spectral energy distribution shows clear infrared excess, indicating the presence of circumstellar dust. The abundance pattern of HE~1153$-$0518 suggests enrichment by early nucleosynthetic processes, such as faint core-collapse supernovae with mixing and fallback, while the possibility of binary interaction and subsequent internal mixing cannot be ruled out. The discovery and detailed study of HE~1153$-$0518 add an important object to the small population of high-A(C) CEMP-no stars and provide valuable constraints on early chemical enrichment pathways and the nature of the first generations of stars.

\end{abstract}

\begin{keywords}
Stars: Individual [HE~0401$-$0138, HE~1153$-$0518, HE~1246$-$1344]; \,
 Stars: Abundances; \,Stars:  AGB; \, Stars: Nucleosynthesis
\end{keywords}

\section{Introduction}

Extremely metal-poor (EMP) stars, with metallicities [Fe/H]~$<$~--3.0 \citep{beers2005discovery}, are among the oldest known objects in the universe and serve as critical indicators of early Galactic chemical enrichment. Their chemical compositions provide a valuable understanding of the nature of the first stars and the nucleosynthetic processes that took place in the early universe \citep{Frebel_&_Norris_2015}. Driven by large-scale surveys and dedicated follow-up studies \citep{beers2005discovery, Frebel_review_2018}, the discovery and study of EMP stars have made significant progress over the past few decades. EMP stars are especially important as they retain the chemical signatures of the early universe, having formed from gas that had been only minutely polluted by prior star formation episodes. Among EMP stars, carbon-enhanced metal-poor (CEMP) stars, characterised by [C/Fe]~$>$~+0.7 \citep{aoki2007carbon,Goswami_et_al_1_2021}, represent a significant fraction \citep{Yoon_et_al_2016}. The CEMP-no subclass, where neutron-capture elements are not enhanced, is thought to have formed from material enriched by the earliest supernovae or by asymptotic giant branch (AGB) stars with minimal $s$-process contributions \citep{Norris_et_al_2013}.

As the sample of CEMP stars increased, several studies began to explore their positions in absolute carbon abundance A(C) and [Fe/H] space. The apparent bimodality in A(C) at different metallicities was first noted by \citep{Rossi_et_al_2005}. \citet{Spite_et_al_2013} examined literature abundance data for about 50 CEMP main-sequence turnoff and dwarf stars, including both CEMP-$s$ and CEMP-no stars, and reported a clear bimodality in A(C), with stars at [Fe/H]~$>$~--3.0, dominated by CEMP-$s$ stars, forming a high-carbon plateau at A(C)~$\approx$~8.25, while stars at [Fe/H]~$<$~--3.4, consisting exclusively of CEMP-no stars, occupied a lower region around A(C)~$\approx$~6.5. \citet{Bonifacio_2015} confirmed and extended this result using a larger sample, finding a similar separation of CEMP-$s$ and CEMP-no stars in A(C)--[Fe/H] space.

Building on these results and using a much larger and more homogenous dataset, \citet{Yoon_et_al_2016} identified three morphological groups in the A(C)--[Fe/H] plane. In this scheme, Group I is dominated by CEMP-$s$ and CEMP-$r/s$ stars with high A(C), while Groups II and III consist primarily of CEMP-no stars. Group II CEMP-no stars show relatively low A(C) and exhibit a clear trend with metallicity, whereas Group III CEMP-no stars display higher A(C) values and do not show any clear metallicity dependence. They also noted the presence of a small number of CEMP-no stars with high A(C) occupying the Group I region, but did not provide a detailed explanation for their origin. Subsequently, \citet{Yoon_et_al_2019} studied the group morphology of CEMP stars in both the Galactic halo and satellite dwarf galaxies and showed that Group I and Group III CEMP-no stars coexist in A(C)--[Fe/H] and A(C)--A(Ba) space. From this behaviour, they suggested that these stars may share a common nucleosynthetic origin, highlighting the complexity of carbon production and early chemical enrichment events.

It is crucial to understand the formation and evolution of EMP and CEMP-no stars in order to reconstruct the early chemical history of the Galaxy and  to constrain the nature of their  progenitor systems \citep{Hansen_et_al_2014}. 
In a recent chemodynamical study of two extremely metal-poor CEMP-no stars HE~1243$-$2408 and HE~0038$-$0345 with [Fe/H] $\sim$ $-$3.05 and $-$2.92 respectively, \citet{Shejeelammal_&_Goswami_2024} demonstrated   that the objects were likely accreted from dwarf spheroidal (dSph) satellite galaxies, thereby supporting hierarchical  assembly scenario of the Galaxy. A detailed analysis of their abundance patterns further indicates that both stars belong to Group~II CEMP-no stars, which exhibit a well-defined trend with metallicity.  
Similarly, a chemical abundance study of the extremely metal-poor stars HE~2148$-$2039 and HE~2155$-$2043 by \citet{Purandardas_&_Goswami_2021}  revealed that HE~2148$-$2039, with $A(\mathrm{C}) < 6.28$, is a Group~II CEMP-no star, whereas HE~2155$-$2043, with $A(\mathrm{C}) \sim 7.05$ and [Fe/H] $< -3.3$, belongs to Group~III. In contrast to Group~II stars, Group~III CEMP-no stars do not exhibit a clear dependence on metallicity. These distinct chemical characteristics provide important constraints on the origin and formation pathways of CEMP-no stars. The discovery and detailed abundance analysis of additional CEMP-no stars are therefore essential for refining models of the first stars and for improving our understanding of Galactic chemical evolution.

In this study, we present a detailed high-resolution spectroscopic analysis of three extremely metal-poor (EMP) stars: HE~0401$-$0138, HE~1153$-$0518, and HE~1246$-$1344. Our results indicate that one of these objects, HE~1153$-$0518, is a CEMP-no star, thereby adding to the growing sample of such stars. This new identification contributes to the expanding database of CEMP-no objects and provides an additional probe of early nucleosynthetic processes. The present analysis is expected to support ongoing efforts to constrain the nature of the first stars and to improve our understanding of the chemical evolution of the early Galaxy.
Section~\ref{sec:previous_studies} presents a review of the previous studies on the programme stars available in the literature. The spectroscopic data used in the present work are described in Section~\ref{sec:source_of_spectra}. In Section~\ref{sec:atmospheric_param}, we explain the methods adopted to determine the radial velocities and stellar atmospheric parameters. Section~\ref{sec:abundance_analysis} outlines the abundance analysis and presents the resulting chemical abundances. A detailed discussion on the chemical peculiarities of the programme stars and their possible formation channels is given in Section~\ref{sec:discussion}. Finally, Section~\ref{sec:conclusion} summarizes the main conclusions of this study.

{\footnotesize
\begin{table*}
\centering
\caption{ \bf{Basic data for the programme stars.}}
\label{tab:basicdata}
\scalebox{0.98}{
\begin{tabular}{lcccccccccccc}
\hline
Star Name    & RA$(2000)$  & Dec.$(2000)$ & B     & V     & J     & H     & K     & Exposure  & Date of Obs. & Spectrograph & Resolving     \\
             &             &              &       &       &       &       &       & (seconds) &              & used         & Power\\
\hline

HE~0401$-$0138& 04 03 49.88 &--01 30 02.45 & 14.72 & 13.84 & 11.95 & 11.46 & 11.30 & 1800 & 27-02-2005 & HDS/SUBARU& 50~000 \\
HE~1153$-$0518& 11 55 54.74 &--05 34 48.14 & 16.50 & 15.07 & 13.25 & 12.71 & 12.56 & 1200 & 28-02-2005 & HDS/SUBARU& 50~000 \\
HE~1246$-$1344& 12 49 20.25 &--14 00 41.59 &  -    & 14.39 & 12.70 & 12.22 & 12.05 & 2700 & 01-03-2005 & HDS/SUBARU& 50~000 \\
\hline
\end{tabular}}
\end{table*}
}

\section{Previous studies on the programme stars}
\label{sec:previous_studies}
\paragraph*{}
\textbf{HE~0401$-$0138 \& HE~1246$-$1344} \\
\citet{barklem2005} presented analysis of ``snapshot" spectra of HE~0401$-$0138 and HE~1246$-$1344 among 253 metal-poor halo stars. The spectra were obtained using VLT/UVES in the wavelength range 3760--4980 {\rm \AA} with R$\sim$20 000. As an initial guess of the atmospheric parameters,  \citet{barklem2005} have considered photometric T$_{eff}$, metallicity estimates from the calibration of the Ca II K line along with B-V colour, log $g$ from log $g$--T$_{eff}$ correlation \citep{Honda_et_al.2004}, microturbulent velocity ($\xi$) = 1.8 km s$^{-1}$, and V$_{macro}$ = 1.5 km s$^{-1}$. The parameters were then refined using an automated analysis based on the Spectroscopy Made Easy (SME) package by \citep{Valenti_&_Piskunov_1996}. \citet{barklem2005} derived the abundances of eleven light elements from C through Ni and two heavy elements, Sr and Ba, for both HE~0401$-$0138 and HE~1246$-$1344. The abundance of Y is also reported for HE~0401$-$0138. \citet{zhang_et_al.2011} adopted the atmospheric parameters from \citet{barklem2005} to derive the abundances of C and Si for HE~0401$-$0138 and HE~1246$-$1344. \citet{Ren_et_al.2012} derived the upper limit of [Th/Fe] in 77 metal-poor stars, including HE~1246$-$1344.

\paragraph*{}
\textbf{HE~1153$-$0518} \\

Atmospheric parameters and elemental abundances for HE~1153$-$0518 based on high-resolution spectra ($R \gtrsim 20{,}000$) have not been previously reported in the literature. In this work, we present the first detailed abundance analysis of this object using high-resolution ($R \sim 50{,}000$) spectroscopy.
  
 The \citet{gaiaDR3_vallenari_et_al_2023} report an effective temperature of $T_{\rm eff} = 4553.3\,\mathrm{K}$ for this star, which is slightly lower than our estimate. However, their derived surface gravity ($\log g = 2.18$) and metallicity differ significantly from our results.While our analysis identifies HE~1153$-$0518 as an extremely metal-poor star with [Fe/H] $= -3.08$, \textit{Gaia} Collaboration et al.~(2023) report a substantially higher value of [Fe/H] $= +0.516$. The large discrepancy in [Fe/H] is likely due to the strong impact of molecular carbon bands on the low-resolution Gaia spectra used by the GSP-Phot pipeline \citep{Lucey_et_al_2023}. Our measured radial velocity of $+72.39\,\mathrm{km\,s^{-1}}$ is marginally lower than the value reported in \textit{Gaia} early DR3  $+76.48 \pm 7.62\,\mathrm{km\,s^{-1}}$ \citep{gaiaDR3_vallenari_et_al_2023}. These significant discrepancies in the derived stellar parameters motivated a more detailed investigation of the chemical properties of this star.

\section{Source of spectra}
\label{sec:source_of_spectra}

\begin{figure}
	\centering
	\includegraphics[height=7cm,width=8.5cm]{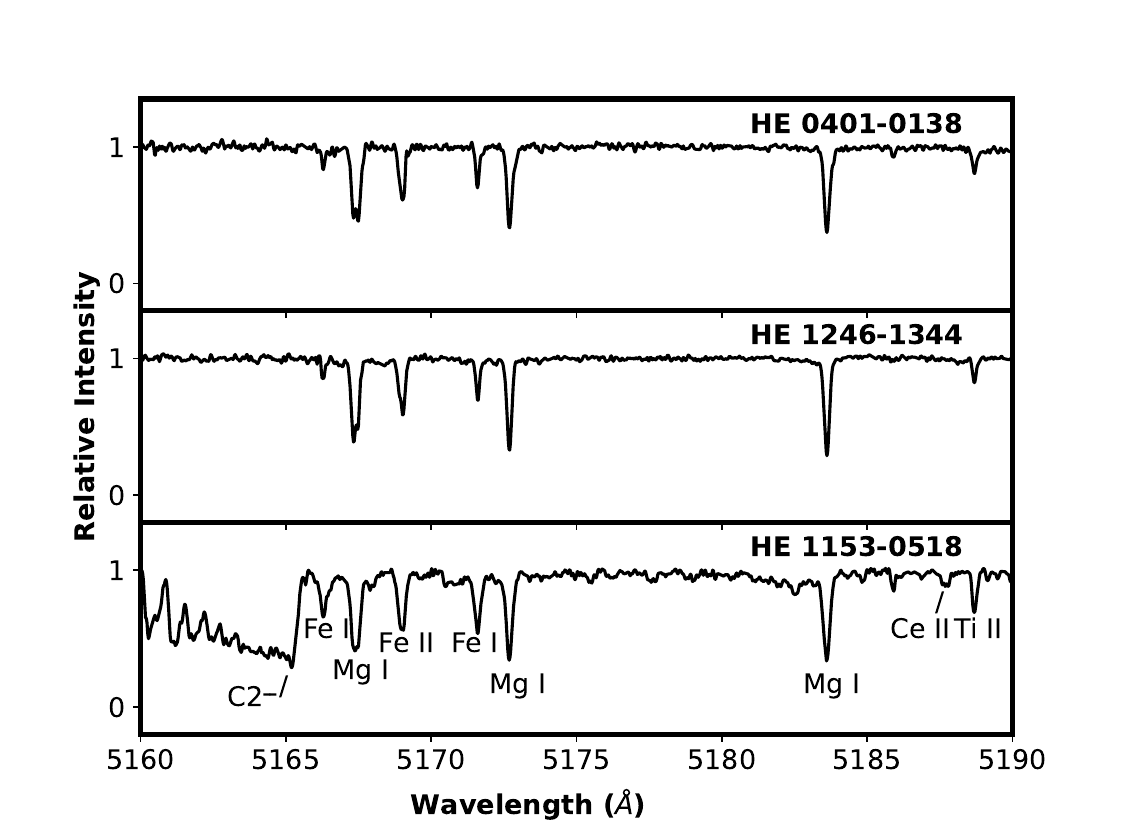}
	\caption{ Sample spectra of the programme stars in the  wavelength 
		region 5160 to 5190 {\bf  {\rm \AA}}.}
\label{fig:sample_spec}
\end{figure}

We have obtained high-resolution (R~$\sim$~50,000) spectra of the stars HE~0401$-$0138, HE~1153$-$0518, and HE~1246$-$1344 from the SUBARU archive\footnote[1]{\url{http://jvo.nao.ac.jp/portal}}. The spectra were acquired using the High Dispersion Spectrograph (HDS) \citep{Noguchi_et_al_2002} mounted to the 8.2-m Subaru Telescope of the National Astronomical Observatory of Japan (NAOJ) at the Mauna Kea Observatory in Hawaii. The wavelength range of the spectra is roughly 4020–6775 {\rm \AA}, with a gap of about 100 {\rm \AA} (between 5340 and 5440 {\rm \AA}) caused by the physical separation of the CCD detectors. The detector consists of two 4100~$\times$~2048 EEV CCDs combined, each having 13.5~$\mu$ pixels. We used the {\it{continuum}} task in the Image Reduction and Analysis Facility (IRAF) to continuum-fit the spectra. Figure~\ref{fig:sample_spec} displays the sample spectra of the programme stars in the 5160--5190 {\rm \AA} wavelength range. Table~\ref{tab:basicdata} provides the basic information about the programme stars.

\section{Radial velocities and atmospheric parameters of the programme stars}
\label{sec:atmospheric_param}
\subsection{Radial Velocities}
\label{sec:radial_velocity}

To determine the radial velocities of the programme stars, we measured the shifts in the wavelengths of several clean and unblended lines in their spectra. We used the Arcturus spectrum \citep{hinkle_et_al_2000} as a template for the laboratory wavelength of the rest frame to ensure homogeneity in the analysis. Arcturus was chosen because it belongs to the giant class and has a temperature comparable to the stars under study. After accounting for heliocentric motion, the calculated mean radial velocities are shown in Table~\ref{tab:radial_velocity_table} along with the standard deviation from the mean values. We also included literature values for comparison. All the three programme stars show variability in radial velocity.

 {\footnotesize
\begin{table}
\centering
\caption{\bf{Radial Velocities of the programme stars.}}
\label{tab:radial_velocity_table} 
\scalebox{0.85}{
\begin{tabular}{lcccc}
\hline
Star Name              & Date of     & V$_{r}$                 & V$_{r}$                & V$_{r}$ \\
                       & Observation & ($km s^{-1}$)           & ($km s^{-1}$)          & ($km s^{-1}$) \\
                       &             & (our estimates)         & \textit{a}             & \textit{b} \\
\hline
HE~0401$-$0138          & 27-02-2005  & +70.02$\pm$0.22         & +137.4                 & +112.27  \\
HE~1153$-$0518          & 28-02-2005  & +72.39$\pm$0.95         &   -                    & +76.48$\pm$7.62 \\
HE~1246$-$1344          & 01-03-2005  & +67.27$\pm$0.41         & +47.90                 & +47.28$\pm$7.61 \\
\hline
\end{tabular}}

\textit{a} -- \citet{barklem2005}; \textit{b} -- \citet{gaiaDR3_vallenari_et_al_2023}\\

\end{table}
}

\subsection{Photometric temperatures}
\label{sec:photometric_temp}

Based on the infrared flux method (IRFM) colour-temperature calibrations, we have calculated the photometric temperatures of the programme stars using broadband colours, optical, and IR. We used the Two Micron All Sky Survey (2MASS) photometric catalogue \citep{2MASS_2003} to get the photometric magnitudes for J, H, and K. We calculated the photometric temperatures for the programme stars using the J--H and V--K colours at various assumed metallicity values, as described in our earlier works \citep{Goswami2006, Goswami2016, Goswami_et_al_1_2021}. Table~\ref{tab:photometric_temp} displays the resulting photometric temperatures. Since T$_{eff}$(J--K) is independent of metallicity \citep{Alonso1996, alonso1999effective}, we used this temperature as an initial guess to choose the correct model atmosphere through an iterative process.

\subsection{Atmospheric parameters from SED fitting}
Spectral Energy distributions (SEDs) can be used to determine stellar parameters and to examine infrared and ultraviolet excess in stars. We have used the Virtual Observatory SED Analyzer (VOSA; \citet{VOSA_2008}) tool to construct and analyse the SEDs. SEDs are constructed from the observed photometric data points spanning a large wavelength range and then fitted with theoretical models to acquire the required stellar parameters. A detailed discussion of the procedure can be found in \citet{Rani_et_al_2021}.

\begin{figure}
	\centering
	\includegraphics[height=5.7cm,width=8.5cm]{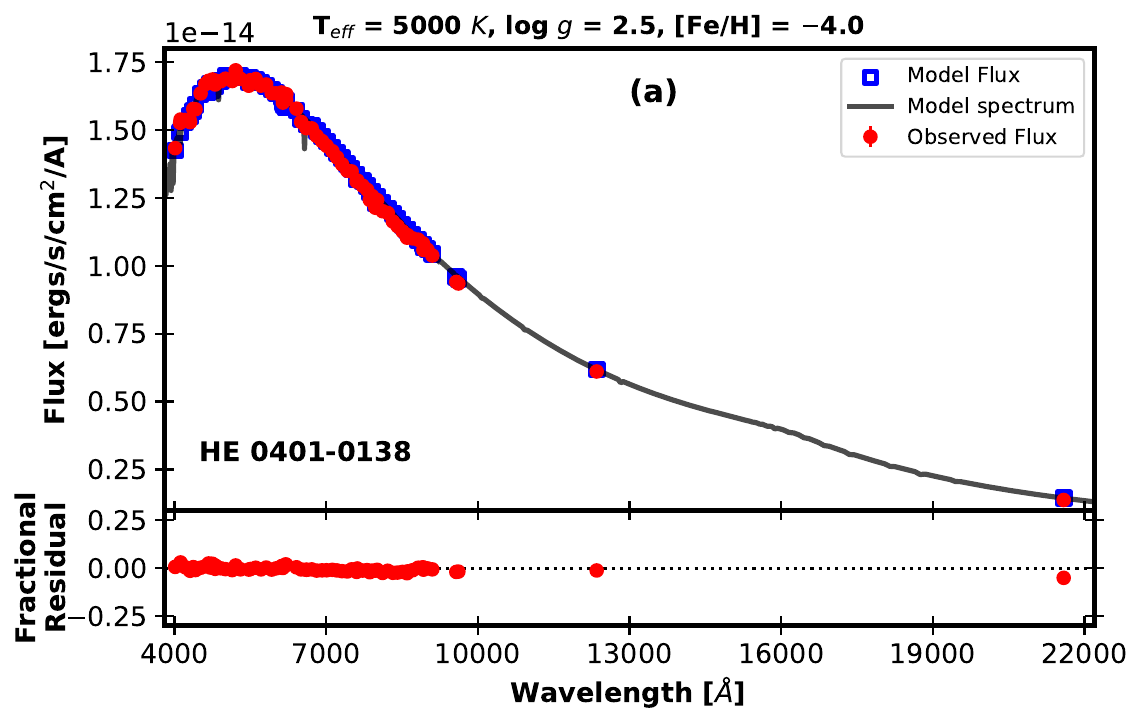}
        \includegraphics[height=5.7cm,width=8.5cm]{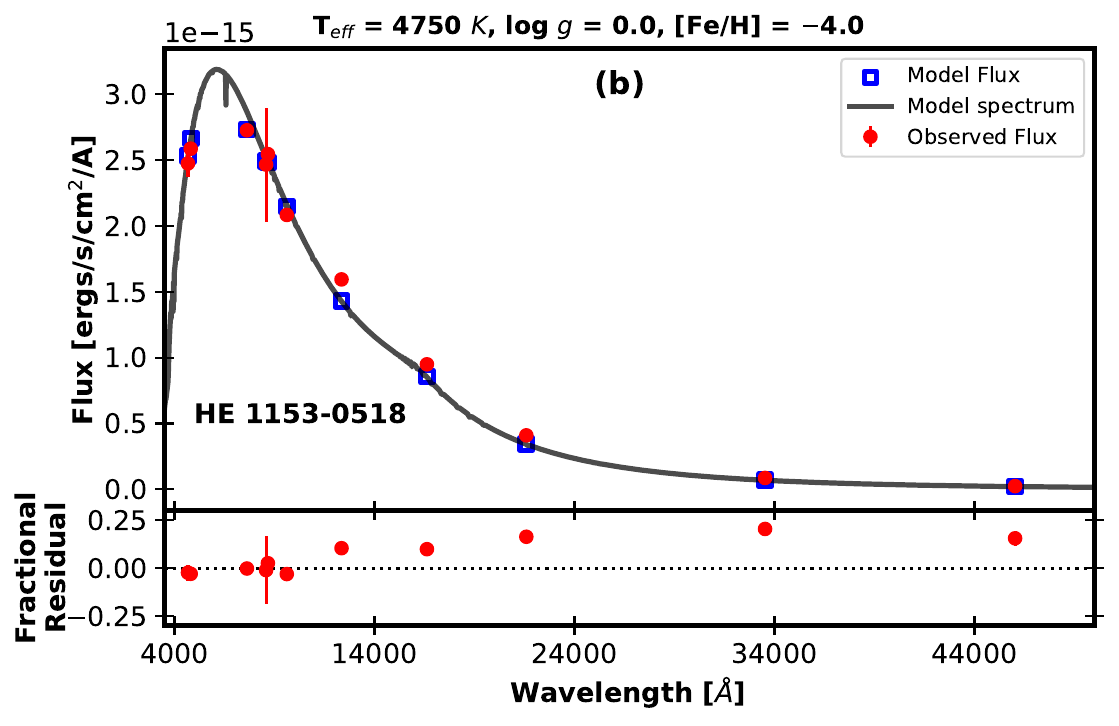}
        \includegraphics[height=5.7cm,width=8.5cm]{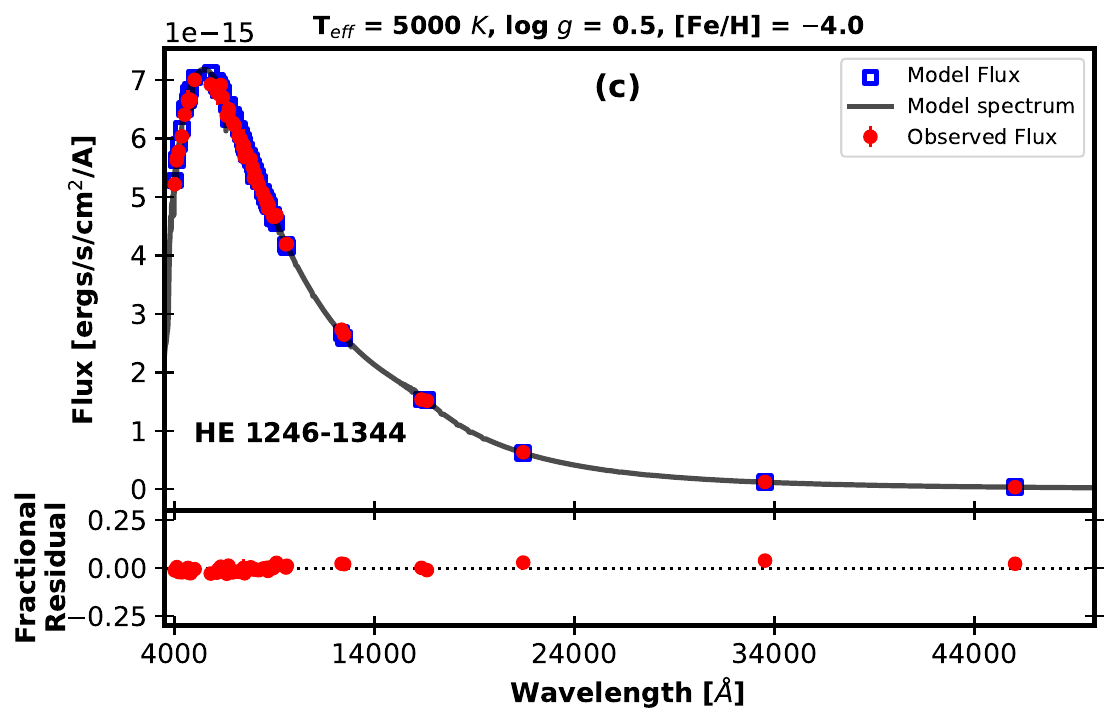}
	\caption{ Spectral energy distribution (SED) fits for the programme stars. The red filled circles with error bars represent the observed fluxes at different wavelengths, while the blue open squares denote the corresponding model fluxes. The black solid curves show the best-fitting Kurucz model spectra. }
\label{fig:SED}
\end{figure}

In order to generate the SEDs, we have used the photometric data from VO catalogues (included in VOSA) and the Kurucz atmospheric models \citep{Kurucz_model_2003} corresponding to [$\alpha$/Fe] = +0.4. The visual extinction (A$_{V}$) for the objects has been 
estimated using the reddening maps provided by \citet{Schlafly_&_Finkbeiner_2011}. For the object HE~0401$-$0138, we have used 73 photometric points covering a wavelength range 4010 -- 21590 {\rm \AA}. Atmospheric model with T$_{eff}$ = 5000 K, log~$g$ = 2.5 and [Fe/H] = --4.0 gave the best fit to the SED (Figure~\ref{fig:SED}(a)) of the object. For HE~1153$-$0518, we have used 11 photometric points covering a wavelength range 4671 -- 46028 {\rm \AA}. Atmospheric model with T$_{eff}$ = 4750 K, log~$g$ = 0.0 and [Fe/H] = --4.0 gave the best fit to the SED (Figure~\ref{fig:SED}(b)) of the object. For HE~1246$-$1344, we have used 61 photometric points covering a wavelength range 4010 -- 46028 {\rm \AA}. Atmospheric model with T$_{eff}$ = 5000 K, log~$g$ = 0.5 and [Fe/H] = --4.0 gave the best fit to the SED (Figure~\ref{fig:SED}(c)) of the object. 

The bottom panels of each subfigure of Figure~\ref{fig:SED} show fractional residuals, that is, the ratio of the difference between the observed flux and model flux to that of the observed flux. The fractional residuals for the objects HE~0401$-$0138 and HE~1246$-$1344 are close to zero for all data points, which indicates that the model spectra represent the SEDs quite well. 
However, the fractional residuals of HE~1153$-$0518 exhibit systematic positive deviations from zero at wavelengths longer than approximately 2~$\mu$m. Although the available photometric coverage is limited to only 11 data points, such an infrared excess is not seen in the other two programme stars, whose SEDs are well reproduced by the models over the entire wavelength range. We therefore interpret this deviation as possible evidence for circumstellar dust emission. Nevertheless, a more robust confirmation would require additional mid-infrared photometric observations. The potential astrophysical implications of this excess are discussed in Section~\ref{sec:discussion}.

{\footnotesize
\begin{table*}
\caption{\bf{Temperatures from  photometry and SEDs. }}
	\label{tab:photometric_temp}
\scalebox{1.0}{
\begin{tabular}{lllllllllllll}
\hline
Star Name         & $T_{eff}$& $T_{eff}$& $T_{eff}$& $T_{eff}$& $T_{eff}$& $T_{eff}$& T$_{eff}$ & SED     & Spectroscopic \\
                  &          & $(-1.5)$ & $(-2.0)$ &  $(-2.5)$& $(-1.5)$ & $(-2.0)$ & $(-2.5)$  & fitting & estimates   \\
                  &  (J-K)   &   (J-H)  &   (J-H)  &  (J-H)   &   (V-K)  &  (V-K)   &   (V-K)   &         & \\
\hline
HE~0401$-$0138     & 4690     & 4967     & 4943     & 4895     & 4505     & 4504     &  4510     & 5000    & 4830  \\
HE~1153$-$0518     & 4546     & 4706     & 4653     & 4639     & 4527     & 4527     &  4533     & 4750    & 4700  \\
HE~1246$-$1344     & 4687     & 4980     & 4956     & 4908     & 4685     & 4683     &  4687     & 5000    & 4780  \\
\hline
\end{tabular}}

The numbers in the parenthesis below $T_{eff}$ represent the metallicities at which the photometric temperatures are calculated. Temperatures are given in Kelvin.\\
\end{table*}
}

\subsection{Spectroscopic determination of atmospheric parameters}
We have determined the atmospheric parameters of the programme stars using a set of clean and unblended Fe~I and Fe~II lines as thoroughly discussed in \citet{Goswami_et_al_1_2021}. These lines have excitation potentials ranging from 0.0~eV to 6.0~eV. Table~\ref{tab:linelist} presents the Fe lines, along with the corresponding atomic data and equivalent widths, used for the estimation of the stellar atmospheric parameters.

The atmospheric models (without convective overshooting) were selected from the Kurucz grid of model atmospheres{\footnote{\url{http://kurucz.harvard.edu/grids.html}}} through an iterative process. For the analysis of the stars, we used an updated version of the radiative transfer code MOOG \citep{Sneden_1973_MOOG}. To choose an initial model atmosphere for a programme star, we used photometric temperature and log~$g$ for giants as initial estimates. The conventional methods of excitation potential balance, equivalent width balance, and ionisation equilibrium of Fe~I and Fe~II lines were applied to determine T$_{eff}$, $\xi$, and log~$g$, respectively. The derived abundances of Fe~I and Fe~II were used to calculate the metallicity, [Fe/H]. Table~\ref{tab:atm_paracomp} presents the atmospheric parameters of the programme stars derived from our analysis alongside their literature values. We used the parameters derived from the spectroscopic method for further analysis.

\begin{figure}
	\centering
	\includegraphics[height=6.7cm,width=8.5cm]{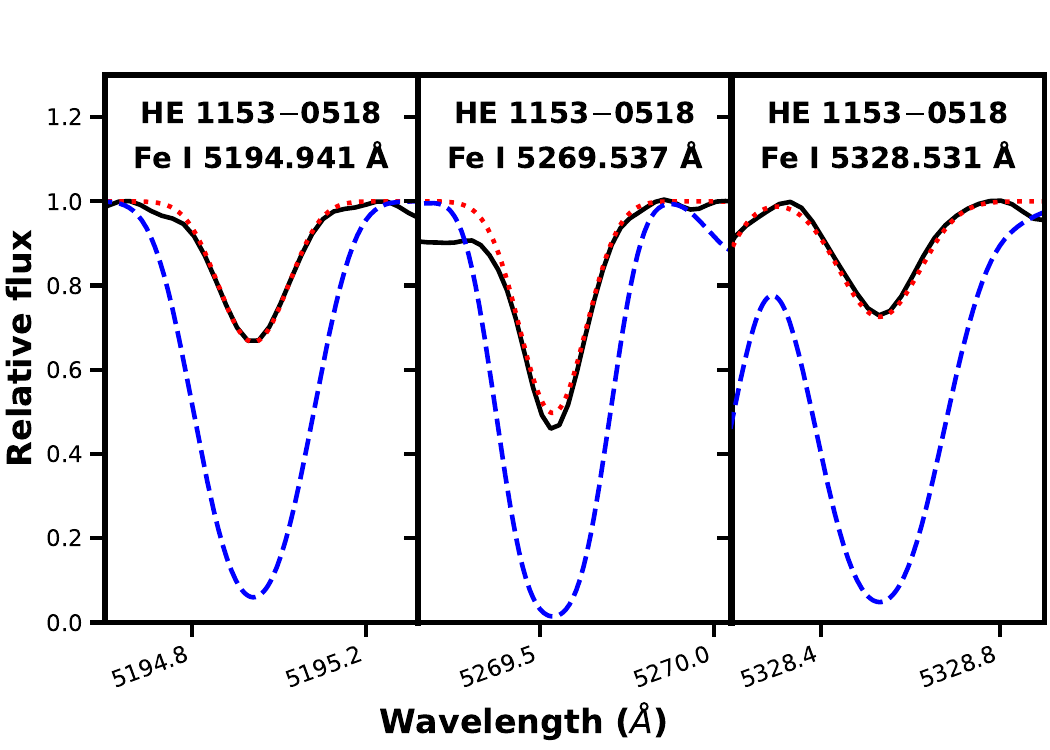}
	\caption{ Spectrum synthesis plots of Fe~I~5194~{\rm \AA}, 5269~{\rm \AA} \& 5328~{\rm \AA} lines for the object HE~1153$-$0518. The solid black lines represent the observed spectra. The red dotted lines show the synthesised spectra computed using the atmospheric parameters derived in this study. The blue dashed lines correspond to the synthesised spectra obtained using the atmospheric parameters reported by \citet{gaiaDR3_vallenari_et_al_2023}.} 
\label{fig:Fe_lines}
\end{figure}

As discussed in Section~\ref{sec:previous_studies}, the atmospheric parameters reported by \citet{gaiaDR3_vallenari_et_al_2023} for HE~1153$-$0518 differ markedly from those derived in this work. To assess the reliability of our parameter estimates, we performed spectral synthesis of selected Fe~I lines in the observed spectrum.  We find that the atmospheric parameters adopted in this study reproduce the observed line profiles satisfactorily,  whereas those reported by \textit{Gaia}  predict significantly  stronger Fe~I absorption features that are not supported by the observations (Figure~\ref{fig:Fe_lines}).

{\footnotesize
\begin{table*}
\centering
\caption{\bf{Equivalent widths (in m\r{A}) of lines used for determination of atmospheric parameters and elemental abundances.}}
\label{tab:linelist} 
\scalebox{1.0}{
\begin{tabular}{cccccccc}
\hline
Wavelength   &Element    &E$_{low}$ &   log gf  &  HE~0401$-$0138  &  HE~1246$-$1344  & HE~1153$-$0518   \\
(\r{A})      &           & (eV)     &           &                 &                 &                 \\
\hline 
4045.81      &  Fe I     &  1.48    &    0.280  &   101.3 (4.10)  &      -          &      -          \\
4198.30      &           &  2.40    & $-$0.719  &        -        &   46.2 (4.16)   &      -          \\
4202.03      &           &  1.48    & $-$0.708  &        -        &   78.6 (3.85)   &      -          \\
4233.60      &           &  2.48    & $-$0.604  &        -        &   33.6 (3.87)   &      -          \\
4250.79      &           &  1.56    & $-$0.710  &    64.9 (3.82)  &      -          &      -          \\
\hline
\end{tabular}}

The numbers in the parenthesis in columns 5 to 7 give the derived abundances from the respective line.\\

\end{table*}}

{\footnotesize
\begin{table*}
\centering
\caption{\bf{Derived atmospheric parameters of our programme stars and literature values. }}
\label{tab:atm_paracomp} 
\scalebox{1.0}{
\begin{tabular}{lccccccccc}
\hline     
Star Name       & T$_{eff}$  &log g  & $\xi    $   & [Fe I/H]          &  [Fe II/H]        & [Fe/H]   & Ref   \\
                &    (K)     & (cgs) & (km s$^{-1}$) &                   &                   &          &       \\
\hline
HE~0401$-$0138  &   4830     &  1.30 &  0.87         & $-$3.46 $\pm$ 0.14 & $-$3.43 $\pm$ 0.06 & $-$3.45  & 1   \\
                &   4866     &  1.76 &  1.45         &     -              &        -           & $-$3.34  & 2   \\
                &   4929     &  -    &  -            &     -              &        -           &   -      & 3   \\
                &   5000     &  2.5  &  -            &     -              &        -           & $-$ 4.0  & 5  \\
\hline
HE~1153$-$0518   &   4700     & 0.20  & 1.47          & $-$3.07$\pm 0.20$   & $-$3.08$\pm 0.16$ & $-$3.08  & 1   \\
                &   4543     & 2.18  &  -            &      -              &      -            &  +0.52   & 4   \\ 
                &   4750     & 0.0   &  -            &      -              &       -           & $-$4.0   & 5   \\
\hline
HE~1246$-$1344  &   4780     &  1.00 &  1.22         & $-$3.51 $\pm$ 0.09 & $-$3.48 $\pm$ 0.01 & $-$3.50  & 1   \\
                &   4853     &  1.65 &  1.84         &    -               &       -            & $-$3.40  & 2   \\
                &   4852     &  -    &   -           &    -               &       -            &   -      & 3   \\
                &   5000     &  0.5  &   -           &    -               &       -            & $-$4.0   & 5   \\              
\hline
\end{tabular}}

References: 1. This work, spectroscopic, 2. \citet{barklem2005}, 3. \citet{gaia2018}, 4. \citet{gaiaDR3_vallenari_et_al_2023}, 5. This work, SED fitting    \\         
\end{table*}
}

\section{Results: Abundance analysis }
\label{sec:abundance_analysis}
We have used the radiative transfer code MOOG 2013 \citep{Sneden_1973_MOOG} for conducting the spectroscopic analysis of the programme stars. MOOG operates under the assumptions of Local Thermodynamic Equilibrium (LTE). For deriving the elemental abundances of the stars, both the equivalent width method and spectrum synthesis method have been employed. The excitation potential and log~$gf$ values for the spectral lines used in the equivalent width method have been taken from the Kurucz database{\footnote{\url{https://lweb.cfa.harvard.edu/amp/ampdata/kurucz23/sekur.html}}} of atomic linelists. The relevant atomic transition data for the lines, along with the measured equivalent widths, are provided in Table~\ref{tab:linelist}. The elemental abundances of the programme stars are presented in Tables~\ref{tab:abundances1}~\&~\ref{tab:abundances_HE1153-0518}.

\subsection{Li, Carbon, Nitrogen, Oxygen}
\label{sec:cno}
\begin{figure}
	\centering
	\includegraphics[height=6.7cm,width=8.5cm]{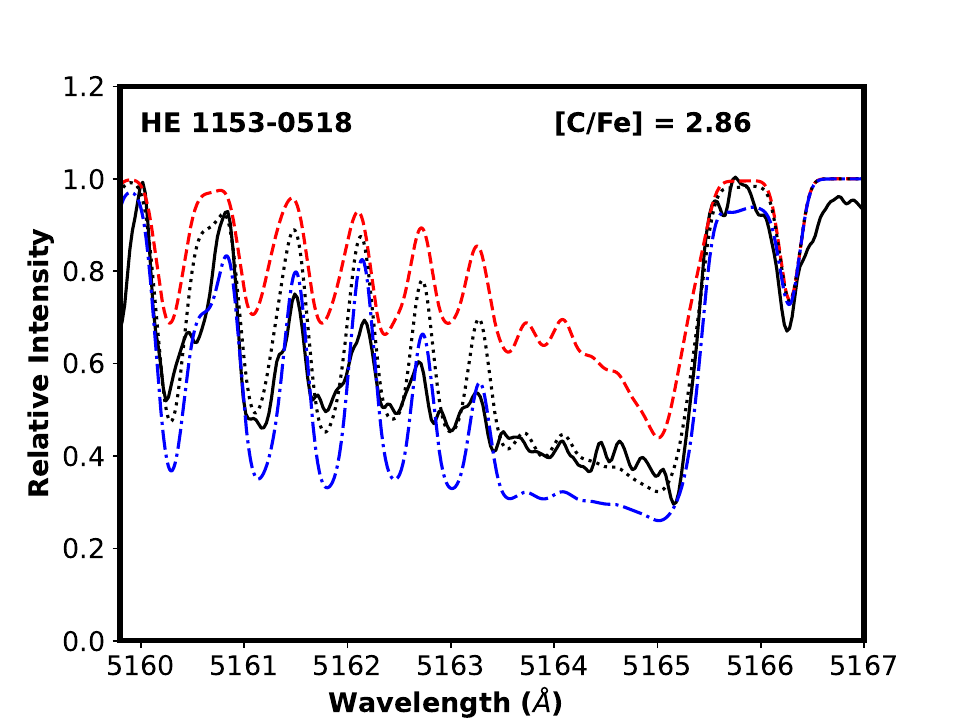}
	\caption{ Spectrum synthesis plot of C$_{2}$ molecular band around 5165 {\rm \AA}. The solid black line represents the observed spectra. The black dotted line shows the synthesised spectra. Other two synthesised spectra are displayed corresponding to $\Delta$[C/Fe] = $-$0.3 and +0.3 with dashed red and dash-dotted blue lines respectively.} 
\label{fig:C2_5165}
\end{figure}

Spectral lines due to lithium and oxygen could not be found in any of the three stars. We have estimated the upper limits of the abundances of Li and O for the programme stars using Li~I~6708 {\rm \AA} and [O~I]~6300 {\rm \AA} lines.  We have derived the abundance of carbon in HE~0401$-$0138 and HE~1246$-$1344 using spectrum synthesis calculations of the CH band at 4310 {\rm \AA}. Other molecular bands due to carbon are absent in these two objects. For the object HE~1153$-$0518, the C$_{2}$ bands at 5165 {\rm \AA} (Figure~\ref{fig:C2_5165}) and 5635 {\rm \AA} have been used to derive the abundance of carbon using spectrum synthesis method. While the abundance of carbon is found to be solar in HE~0401$-$0138 ([C/Fe] = +0.29) and HE~1246$-$1344 ([C/Fe] = +0.02) HE~1153$-$0518 exhibits enhancement of carbon with [C/Fe] = +2.86. CN band at 4215 {\rm \AA} could not be detected in the spectra of any objects. So, we could not determine the abundance of nitrogen in the programme stars. The isotopic ratio of carbon, $^{12}$C/$^{13}$C, could be derived only for HE~1153$-$0518 using the spectrum synthesis calculations of C$_{2}$ band at 4740 {\rm \AA}. This particular band is absent in the other two stars. The linelists of the CH band at 4310 {\rm \AA} and the C$_{2}$ bands at 4740 {\rm \AA}, 5165 {\rm \AA} and 5635 {\rm \AA}  are adopted from the {\textit{linemake}}{\footnote{\textit{linemake} contains laboratory atomic data (transition probabilities, hyperfine and isotopic substructures) published by the Wisconsin Atomic Physics and the Old Dominion Molecular Physics groups. These lists and accompanying line list assembly software have been developed by C. Sneden and are curated by V. Placco at \url{https://github.com/vmplacco/linemake.}}} atomic and molecular line database.

\subsection{Light Elements}
We have derived the abundances of Na, Mg, Ca, Ti, Cr and Ni for all three programme stars using the equivalent width technique. For Ti, spectral lines due to both the available species Ti~I and Ti~II have been used to derive the abundances. Both the species yielded abundance consistent with error bars. Usable lines due to Co could not be detected in the spectrum of the star HE~1153$-$0518. For the other two stars, we could derive the abundance of Co. As Sc and Mn show hyperfine splitting, we used the spectrum synthesis technique to estimate the abundances of Sc and Mn. While Sc~II 4246.822 {\rm \AA} \& Sc~II 4320.732 {\rm \AA} lines are used for HE~0401$-$0138 and HE~1246$-$1344, only Sc~II 4246.822 {\rm \AA} could be used for HE~1153$-$0518. We could not detect any good Mn line in the spectrum of HE~1153$-$0518. For HE~0401$-$0138, we have used Mn~I 4030.753 {\rm \AA} \& Mn~I 4033.062 {\rm \AA} lines to derive the abundance of Mn. For HE~1246$-$1344, we have used Mn~I 4033.062 {\rm \AA} \& Mn~I 4766.418 {\rm \AA} lines to derive the Mn abundance. The hyperfine contributions to the Sc and Mn lines are adopted from \textit{linemake}. 

\subsection{Neutron-capture Elements}
\begin{figure*}
     \begin{center}
\centering
        \subfigure[Light elements]{%
            \label{fig:light}
            \includegraphics[height=8.0cm,width=9.0cm]{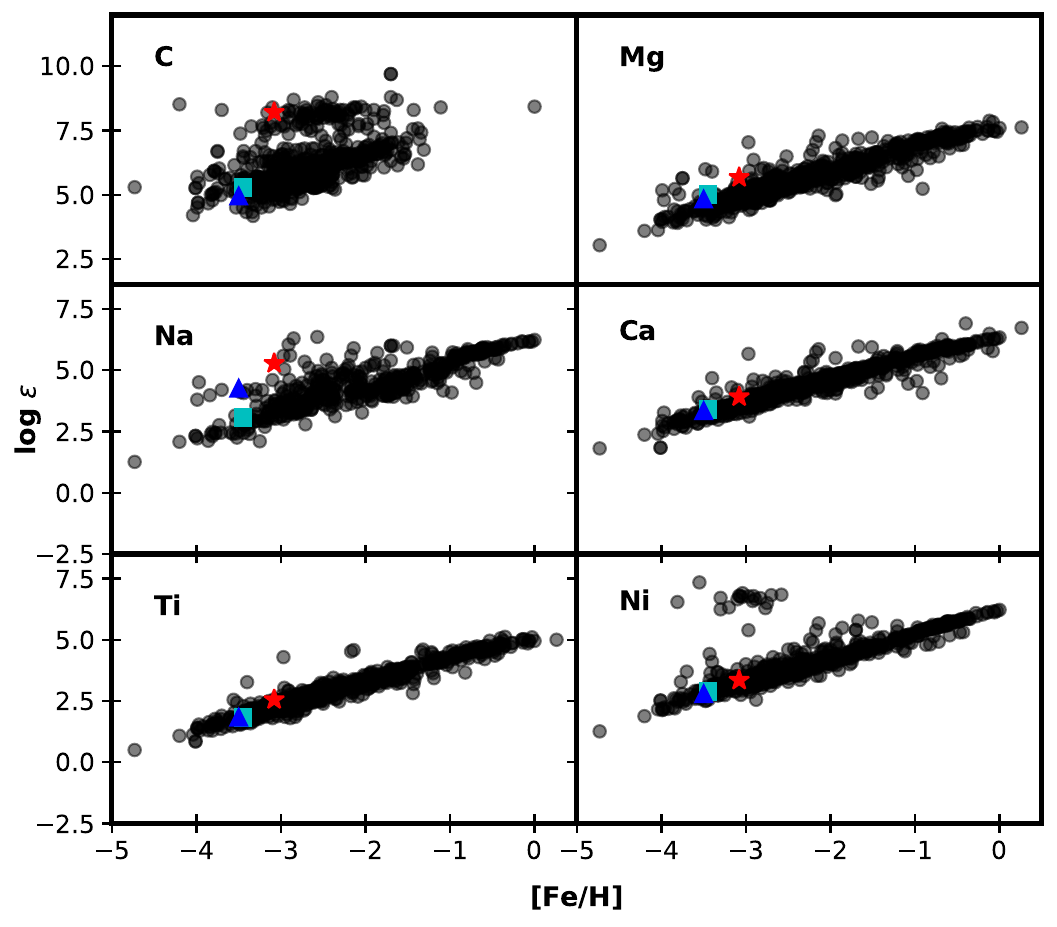}
        }%
        \subfigure[Heavy elements]{%
            \label{fig:heavy}
            \includegraphics[height=8.0cm,width=9.0cm]{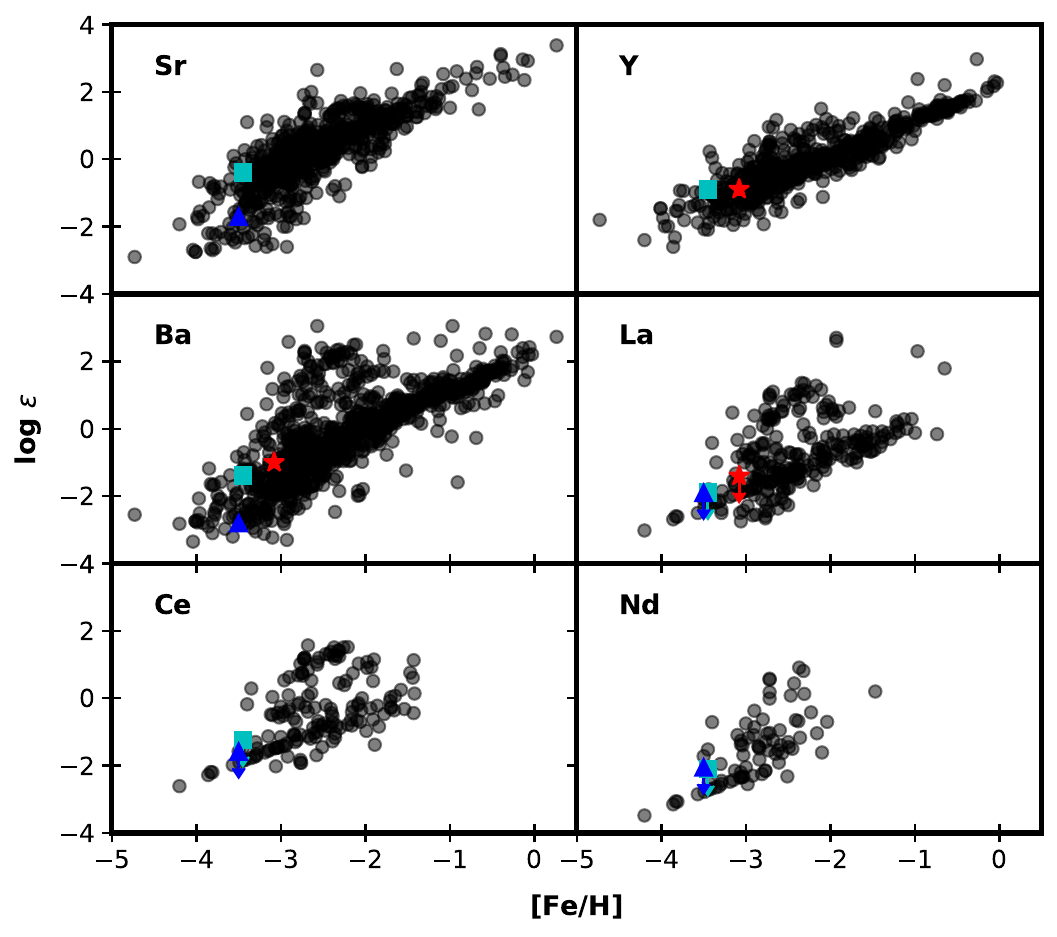}

        }\\ 

\caption{Comparison of the observed abundances of a few light and heavy elements of our programme stars with the literature data of metal-poor stars compiled by \citet{Frebel_2010}. The grey-filled circles represent the literature abundance values of metal-poor stars. Cyan-filled squares represent HE~0401$-$0138, blue-filled triangles represent HE~1246$-$1344 and red-filled stars represent HE~1153$-$0518. The error bars to the abundances of the corresponding elements are not visible in the plots as the marker sizes are larger than the error bars. }%
   \label{fig:elements}
       \end{center}

\end{figure*}

We have derived the abundances of Sr for HE~0401$-$0138 and HE~1246$-$1344. Sr is found to be solar in HE~0401$-$0138 and subsolar in HE~1246$-$1344. We could not find any good line due to Sr for the object HE~1153$-$0518. Abundance of Y has been derived for HE~0401$-$0138 and HE~1153$-$0518 and it is found to be solar in both the objects. Abundance of Y could not be derived for the other object due to the absence of clean and usable lines in the spectrum. The absorption lines used for the equivalent width method can be found in Table~\ref{tab:linelist}. We have used spectrum synthesis calculations to derive the abundance of Ba for all three objects. For HE~0401$-$0138, we have used Ba~II 6141.713 {\rm \AA} and Ba~II 6496.897 {\rm \AA} lines; For HE~1246$-$1344, we have used Ba~II 4554.029 {\rm \AA} line and for HE~1153$-$0518, we have used Ba~II 4554.029 {\rm \AA}, Ba~II 5853.668 {\rm \AA} and Ba~II 6141.713 {\rm \AA} lines. Ba is found to be sub-solar in all three stars. We could not find any lines due to La, Ce, Pr, Nd, Sm and Eu in any of the objects. However, we have estimated the upper limits of these elements in HE~0401$-$0138 and HE~1246$-$1344. We could derive the upper limit of the abundance of La in HE~1153$-$0518. In Figure~\ref{fig:elements}, we have compared the abundances (log~$\epsilon$) of a few light and heavy elements of our programme stars with that of other metal-poor stars compiled by \citet{Frebel_2010} from literature. Figure~\ref{fig:light} \& \ref{fig:heavy} clearly show that all three programme stars exhibit elemental abundances consistent with the Galactic metal-poor stars.

We note that we have not accounted for possible non-local thermodynamic equilibrium (NLTE) and three-dimensional (3D) effects \citep{Asplund_2005, Lind_et_al_2011} in the present analysis. These effects are not expected to alter the qualitative classification of the programme stars or the main interpretations based on their abundance patterns \citep{Yoon_et_al_2016}.

{\footnotesize
\begin{table*}
\caption{\bf{Elemental abundances in HE~0401$-$0138 and HE~1246$-$1344}}
\label{tab:abundances1}
\begin{tabular}{l c c|c c c|c|| c c c|c}
\hline
& \multicolumn{6}{c}{HE~0401$-$0138} & \multicolumn{0}{c}{HE~1246$-$1344}\\
\hline
             & Z  & solar $log{\epsilon}^{(a)}$ & $log{\epsilon}$ &[X/H]& [X/Fe] & [X/Fe]    & $log{\epsilon}$ &[X/H]& [X/Fe] & [X/Fe]\\
	         &    &                         &    (dex)        &     &        & $^{(b)}$  &    (dex)        &     &        & $^{(b)}$      \\
\hline
Li                   & 3  & 1.05 &$<$ 0.80             &$<$ --0.25 &$<$ +3.20 &            &$<$ 0.30             &$<$ --0.75 &$<$ +2.75 & -         \\
C (CH, 4310{\rm \AA})& 6  & 8.43 & 5.27 (syn)          & $-$3.16   & +0.29    &  +0.24       & 4.95 (syn)          & $-$3.48 & +0.02      &  $-$0.06  \\
O                    & 8  & 8.69 &$<$ 6.80 (1)         &$<$ --1.89 &$<$ +1.56 &             &$<$ 6.58 (1)         &$<$--2.11&$<$ +1.39   &  -           \\
Na {\sc i}           & 11 & 6.24 & 3.07$\pm$0.19 (2)   & $-$3.17   & +0.28    &   -         & 4.27$\pm$0.02 (2)   & $-$1.97 & +1.53      &   -       \\
Mg {\sc i}           & 12 & 7.60 & 5.01$\pm$0.12 (3)   & $-$2.59   & +0.86    &  +0.42       & 4.83$\pm$0.20 (2)   & $-$2.77 & +0.73      &  +0.50     \\
Ca {\sc i}           & 20 & 6.34 & 3.41$\pm$0.02 (3)   & $-$2.93   & +0.52    &  +0.35       & 3.35$\pm$0.14 (7)  & $-$2.99 & +0.51      &  +0.39     \\
Sc {\sc ii}$^{*}$    & 21 & 3.15 & $-$0.05$\pm$0.05 (2)& $-$3.20   & +0.25    &  +0.11       & $-$0.10$\pm$0.14 (2)&$-$3.25  & +0.25      &  +0.35     \\
Ti {\sc i}           & 22 & 4.95 & 1.84$\pm$0.09 (3)   & $-$3.11   & +0.34    &  +0.28       & 1.84$\pm$0.05 (2)   & $-$3.11 & +0.39      &  +0.25     \\
Ti {\sc ii}          & 22 & 4.95 & 1.99$\pm$0.23 (6)   & $-$2.96   & +0.49    &  -          & 1.86$\pm$0.10 (12)  & $-$3.09 & +0.41      &   -      \\
Cr {\sc i}           & 24 & 5.64 & 2.14$\pm$0.08 (2)   & $-$3.50   & $-$0.05 & $-$0.46     & 1.87$\pm$0.17 (4)   & $-$3.77   &$-$0.27  &  $-$0.45 \\
Mn {\sc i}$^{*}$     & 25 & 5.43 & 1.27$\pm$0.03 (2)   & $-$4.16   & $-$0.71 & $-$0.80     & 1.12$\pm$0.08 (2)   & $-$4.31   & $-$0.81 &  $-$0.91 \\
Fe {\sc i}           & 26 & 7.50 & 4.04$\pm$0.14 (33)  & $-$3.46   & -       &   -         & 3.99$\pm$0.09 (43)  & $-$3.51   & -       &  -        \\
Fe {\sc ii}          & 26 & 7.50 & 4.07$\pm$0.06 (4)   & $-$3.43   & -       &   -         & 4.02$\pm$0.01 (2)   & $-$3.48   & -       &  -        \\
Co {\sc i}           & 27 & 4.99 & 1.76 (1)            & $-$3.23   & +0.22    &  +0.24       & 1.62 (1)            & $-$3.37   & +0.13    &  +0.12   \\
Ni {\sc i}           & 28 & 6.22 & 2.89 (1)            & $-$3.33   & +0.12    &  +0.02       & 2.80$\pm$0.09 (2)   & $-$3.42   & +0.08    &  $-$0.01 \\
Sr {\sc ii}          & 38 & 2.87 &$-$0.39 (1)          & $-$3.24   & +0.21    & +0.11        &$-$1.70$\pm$0.11 (2) & $-$4.57   &$-$1.07  & $-$1.25\\
Y {\sc ii}           & 39 & 2.21 &$-$0.90 (1)          & $-$3.11   & +0.34    & +0.05        &    -                &   -       &   -     &   -    \\
Ba {\sc ii}$^{*}$    & 56 & 2.18 &$-$1.39$\pm$0.01 (2)& $-$3.57    &$-$0.12  & $-$0.21     &$-$2.80 (1)         & $-$4.98    &$-$1.48  &   -    \\
La {\sc ii}$^{*}$    & 57 & 1.10 & $<$ $-$1.90$\pm$0.10 (2)&$<$ $-$3.00 &$<$ +0.45 &   -         & $<$ $-$1.90$\pm$0.10 (2)& $<$ $-$3.00&$<$ +0.50 &   -    \\
Ce {\sc ii}$^{*}$    & 58 & 1.58 & $<$ $-$1.24$\pm$0.16 (2)&$<$ $-$2.82 &$<$ +0.63 &   -         & $<$ $-$1.58 (1)         & $<$ $-$3.16&$<$ +0.34 &   -    \\
Pr {\sc ii}$^{*}$    & 59 & 0.72 & $<$ $-$2.15$\pm$0.05 (2)&$<$ $-$2.87 &$<$ +0.58 &   -         & $<$ $-$1.82$\pm$0.10 (2)& $<$ $-$2.54&$<$ +0.96 &   -    \\
Nd {\sc ii}$^{*}$    & 60 & 1.42 & $<$ $-$2.10$\pm$0.10 (2)&$<$ $-$3.52 &$<$ --0.07&  -         & $<$ $-$2.05$\pm$0.05 (2)& $<$ $-$3.47&$<$ +0.03 &   -\\
Sm {\sc ii}$^{*}$    & 62 & 0.96 & $<$ $-$2.36 (1)         &$<$ $-$3.32 &$<$ +0.13 &   -         & $<$ $-$1.87$\pm$0.24 (3)& $<$ $-$2.83&$<$ +0.67 &   - \\
Eu {\sc ii}$^{*}$    & 63 & 0.52 & $<$ $-$2.20$\pm$0.10 (2)&$<$ $-$2.72 &$<$ +0.73 &   -         & $<$ $-$2.35$\pm$0.15 (2)& $<$ $-$2.87&$<$ +0.63 &   -         \\
\hline 
\end{tabular}
 
$^{*}$ abundance is derived using spectrum synthesis calculations. The number inside the parenthesis shows the number of lines used for the abundance determination.\\ \textbf{References:} $^{(a)}$ \citet{asplund2009}, $^{(b)}$\citet{barklem2005}.\\
\end{table*}
}

{\footnotesize
\begin{table*}
\caption{\bf{Elemental abundances in HE~1153$-$0518.}}
\label{tab:abundances_HE1153-0518}
\begin{tabular}{l c c|c c c}
\hline
\hline
            & Z  & solar $log{\epsilon}^{(a)}$ & $log{\epsilon}$           &[X/H]    & [X/Fe]      \\
	        &    &                             &    (dex)                  &         &             \\
\hline
Li                          & 3  & 1.05 & $<$ 0.90            &$<$ --0.15  &$<$ +2.93        \\
C (C$_{2}$, 5165 {\rm \AA}) & 6  & 8.43 & 8.20 (syn)          & $-$0.23    & +2.85           \\
C (C$_{2}$, 5635 {\rm \AA}) & 6  & 8.43 & 8.22 (syn)          & $-$0.21    & +2.87           \\
O                           & 8  & 8.69 &$<$ 6.89 (syn)       &$<$ --1.80  &$<$ +1.28        \\
Na {\sc i}                  & 11 & 6.24 & 5.28$\pm$0.11 (2)   & $-$0.96    & +2.12           \\
Mg {\sc i}                  & 12 & 7.60 & 5.57$\pm$0.05 (2)   & $-$2.03    & +1.05           \\
Ca {\sc i}                  & 20 & 6.34 & 3.92 (1)            & $-$2.42    & +0.66           \\
Sc {\sc ii}$^{*}$           & 21 & 3.15 &--0.07 (1)           & $-$3.22    & --0.15         \\ 
Ti {\sc i}                  & 22 & 4.95 & 2.56 (1)            & $-$2.39    & +0.69           \\
Ti {\sc ii}                 & 22 & 4.95 & 2.33$\pm$0.16 (4)   & $-$2.62    & +0.46           \\
Cr {\sc i}                  & 24 & 5.64 & 2.70 (1)            & $-$2.94    & +0.14          \\
Fe {\sc i}                  & 26 & 7.50 & 4.43$\pm$0.20 (16)  & $-$3.07    & -              \\
Fe {\sc ii}                 & 26 & 7.50 & 4.42$\pm$0.16 (2)   & $-$3.08    & -              \\
Ni {\sc i}                  & 28 & 6.22 & 3.35 (1)            & $-$2.87    & +0.21           \\
Y {\sc ii}                  & 39 & 2.21 &$-$0.89 (1)          & $-$3.10    & --0.02           \\
Ba {\sc ii}$^{*}$           & 56 & 2.18 &--1.0$\pm$0.17 (3)   & $-$3.18    & --0.10           \\
La {\sc ii}$^{*}$           & 57 & 1.10 &$<$ --1.40 (1)        &$<$ $-$2.50    &$<$ +0.58           \\
\hline 
                       \\
$^{12}$C/$^{13}$C = 2.0\\
                       \\
\hline
\hline
\end{tabular}

$^{*}$ abundance is derived using spectrum synthesis calculations. The number inside the parenthesis shows the number of lines used for the abundance determination.\\  $^{(a)}$ \citet{asplund2009}.\\
\end{table*}
}

 \begin{table*}
\centering
\hspace{2.0cm}\caption{\bf {Spatial velocity and probability estimates.}}
\scalebox{1.0}{
\begin{tabular}{lccccccc}
\hline
Star Name       & $U_{LSR} (km/s) $  & $V_{LSR} (km/s)$       &$ W_{LSR} (km/s)$        & $V_{spa}$ (km/s)    & $P_{thin} $ & $P_{thick}$& $ P_{halo}$\\
\hline
HE~0401$-$0138  &$-$57.84 $\pm$ 2.87 &$-$256.53 $\pm$ 48.18   & 54.79 $\pm$ 16.90       & 268.61 $\pm$ 42.95  & 0.00        & 0.00       &  1.00 \\

HE~1153$-$0518  &$-$297.30 $\pm$ 157.76 & $-$360.34 $\pm$ 165.75 & $-133.06$ $\pm$ 99.83 & 485.73 $\pm$ 348.21 &  0.00     & 0.00       &  1.00 \\

HE~1246$-$1344  & 27.87 $\pm$ 6.73   & $-$475.47 $\pm$ 271.18 & $-$273.19 $\pm$ 199.48  & 549.08 $\pm$ 332.01 & 0.00        & 0.00       &  1.00 \\
\hline

\label{tab:kinematicresults}    
\end{tabular}
}
\end{table*}

{\footnotesize
\begin{table*}
\centering
\hspace{2.5cm}\caption{ \bf{Classification of the programme stars.}}
\label{tab:classification_3}
\scalebox{1.0}{
\begin{tabular}{lcccccccl}
\hline
Star Name       &  [Fe/H]  &  [C/Fe]  &  [Ba/Fe]  &  [La/Fe]  &  [Eu/Fe]  &  [Ba/Eu]  &  [La/Eu]  &  Classification      \\
                &          &          &           &           &           &           &           &                      \\
\hline
HE~0401$-$0138   &  --3.45  &  +0.29    & --0.12    & $<$ +0.45  &  $<$ +0.73 &$>$ --0.85 &$>$ --0.28 & EMP                  \\
HE~1153$-$0518   &  --3.08  &  +2.86    & --0.10    & $<$ +0.58  &    -      &  -        &  -        & CEMP-no              \\
HE~1246$-$1344   &  --3.50  &  +0.02    & --1.48    & $<$ +0.50  & $<$ +0.63  &$>$ --2.11 &$>$ --0.13 & EMP                  \\
\hline

\end{tabular}}
\end{table*}
}

\subsection{Kinematic Analysis}
 In order to know the galactic population to which the programme stars belong, we performed a kinematic analysis of the stars. We followed the procedure described in \citet{meenakshi2019chIII} and \citet{Goswami_et_al_1_2021}. We calculated the components of space velocities (U$_{LSR}$, V$_{LSR}$, W$_{LSR}$) with respect to local standard of rest (LSR). The values of the parallaxes ($\pi$) and proper motions ($\mu_{\alpha}$, $\mu_{\delta}$) are taken from the Gaia database. Radial velocities of the programme stars are taken from our estimates. V$_{spa}$ (= $\sqrt{U_{LSR}^{2} + V_{LSR}^{2} + W_{LSR}^{2}}$) gives the total space velocity. We have calculated the probabilities of the programme stars being in the thin-disk (P$_{thin}$), thick-disk (P$_{thick}$) or halo (P$_{halo}$) population using the procedures given in \citet{reddy2006}, \citep{bensby2003, bensby2004} and \citet{mishenina2004}. 

For HE~0401$-$0138 and HE~1246$-$1344, the parallaxes from Gaia were used directly. However, for HE~1153$-$0518, the Gaia parallax is negative and unreliable. Therefore, we estimated the distance of this star using the method and equations discussed in \citet{meenakshi2019chIII}. We assumed a stellar mass of 0.8 M$_{\odot}$ and used the value of $A_V$ from \citet{Schlafly_&_Finkbeiner_2011}. We adopted a bolometric correction of BC = $-$0.48, a value supported by the calibration of \citet{alonso1999effective} for cool giant stars. Using the estimated distance, we calculated the kinematic parameters for HE~1153$-$0518 in the same way as for the other two stars.
 
 The values of U$_{LSR}$, V$_{LSR}$, W$_{LSR}$, V$_{spa}$, P$_{thin}$, P$_{thick}$ and P$_{halo}$ for all three programme stars are presented in Table~\ref{tab:kinematicresults}.

\section{Discussions}
\label{sec:discussion}

In this section, we discuss the implications of the derived abundance patterns of the programme stars. We aim to understand the possible formation scenarios of these stars using well-established abundance diagnostics and comparisons with previous studies.

\subsection{Classification of the programme stars}

Based on the standard classification scheme of extremely metal-poor (EMP) stars ([Fe/H]~$<$~--3.0; \citet{beers2005discovery, Frebel_review_2018}), all three programme stars are classified as EMP stars. Among them, HE~0401$-$0138 and HE~1246$-$1344 exhibit no carbon enhancement and show abundance patterns typical of normal EMP halo stars, characterised by mild $\alpha$-element enhancement and low neutron-capture element abundances  (Figure~\ref{fig:elements}). \citet{barklem2005} also studied these two objects and reported the abundances of a few elements. For the elements common in both the studies our derived abundances closely match with that of \citet{barklem2005}. In addition to the elements reported by \citet{barklem2005}, we derived the abundance of sodium for the objects. We also derived the upper limits of the abundances of Li, O, La, Ce, Nd, Pr, Sm and Eu. The probability estimates support the inclusion of these two objects in the halo population.

The moderate enhancement of the $\alpha$-elements ([(Mg, Ca, Ti)/Fe] $\sim$ +0.34 to +0.86) in the EMP stars HE~0401$-$0138 and HE~1246$-$1344 is consistent with enrichment dominated by core-collapse supernovae of massive stars, as commonly observed in EMP stars (e.g. \citet{Cayrel_et_al_2004}). The low abundances of Fe-peak elements such as Mn further support this picture, as Mn is known to be underproduced in low-metallicity supernovae \citep{Cayrel_et_al_2004, Kobayashi_et_al_2006}. We note that HE~1246–1344 shows a higher sodium abundance compared to HE~0401–-0138. Such star-to-star variations in light odd-Z elements, particularly sodium, are commonly observed among EMP halo stars and do not necessarily imply different formation channels. The sodium yield in core-collapse supernovae is known to be sensitive to progenitor mass, explosion energy, and fallback efficiency, as well as to the degree of mixing in the ejecta (e.g. \citet{Kobayashi_et_al_2006, Nomoto_et_al_2013}).

HE~1153$-$0518, on the other hand, shows very large carbon enhancement ([C/Fe]~=~+2.86) and low abundances of neutron-capture elements ([(Y, Ba)/Fe]~$<$~0.0). According to the classification criteria of CEMP stars \citep{beers2005discovery, Frebel_review_2018, hansen2019abundances}, HE~1153$-$0518 is a bona fide CEMP-no star. Table~\ref{tab:classification_3} presents the ratios of a few key elements and the classification of the programmes stars. In the following subsections we will discuss the findings about HE~1153$-$0518 in detail.

\subsection{Position of HE~1153$-$0518 on the Yoon–Beers diagram and its Group~IV membership}

\begin{figure}
	\centering
	\includegraphics[height=6.7cm,width=8.5cm]{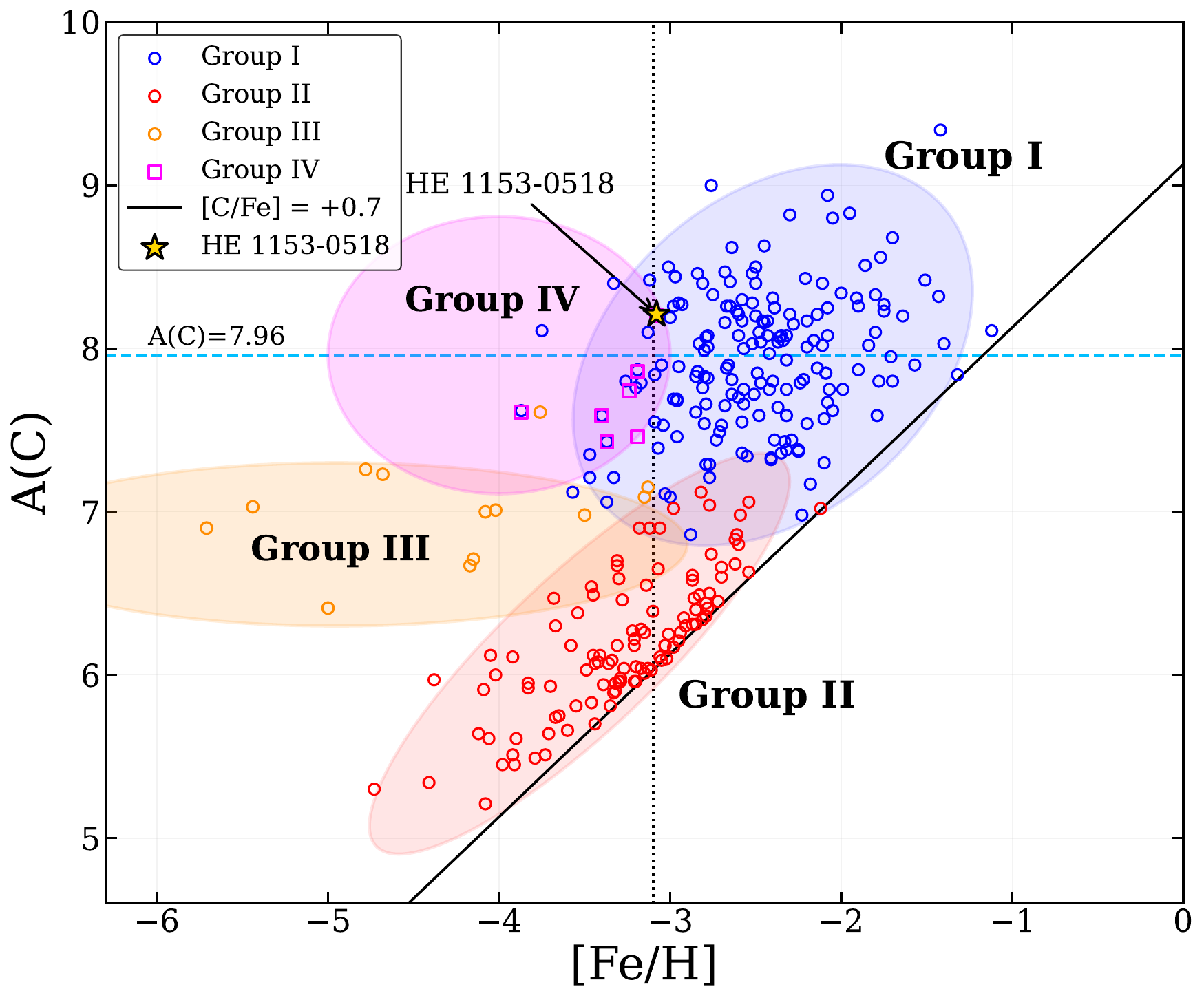}
	\caption{ Yoon--Beers diagram showing the absolute carbon abundance, A(C), as a function of metallicity, [Fe/H], for carbon-enhanced metal-poor (CEMP) stars. Blue circles, red circles, orange circles, and magenta squares represent Groups~I, II, III, and IV, respectively. The large yellow star marks HE~1153$-$0518 from this work. The coloured ellipses show the approximate locations of the different groups, following \citet{Yoon_et_al_2016} for Groups~I--III and \citet{Lee_et_al_2025} for Group~IV. The sample includes the full compilation from \citet{Yoon_et_al_2016} together with the stars explicitly listed in \citet{Lee_et_al_2025} and \citet{Hong_et_al_2026}. The cyan dashed horizontal line at A(C) = 7.96 marks the high-carbon peak identified by \citet{Yoon_et_al_2016}. The vertical dotted line at [Fe/H] = $-$3.1 approximately marks the metallicity region associated with Group~IV. The solid black line represents the CEMP classification boundary corresponding to [C/Fe] = +0.7.} 
\label{fig:YB_diagram}
\end{figure}

Figure~\ref{fig:YB_diagram} shows the location of HE~1153$-$0518 in the Yoon--Beers A(C)$-$[Fe/H] diagram. The star has A(C) $\approx$ 8.2 at [Fe/H] = $-$3.08 and therefore lies in the high-A(C) region.

Recently, \citet{Lee_et_al_2025} identified a new morphological group of CEMP stars, which they designated as Group~IV. These stars are characterised by high absolute carbon abundances (A(C) $>$ 7.39) and low metallicities ([Fe/H] $\leq$ $-$3.1), but they lack significant neutron-capture element enhancements and are therefore classified as CEMP-no stars. HE~1153$-$0518 satisfies all these criteria and thus belongs to the newly proposed Group~IV.

\citet{Lee_et_al_2025} suggested that Group~IV stars may have formed in relatively carbon-normal metal-poor environments and later acquired additional carbon through mass transfer from an AGB companion in a binary system. The high binary fraction ($\sim$30$\%$) they reported supports this scenario. The properties of HE~1153$-$0518, in particular its extreme sodium enhancement ([Na/Fe] = +2.12), very low carbon isotopic ratio ($^{12}C/^{13}C$ = 2.0), and the presence of circumstellar dust, provide further observational constraints on the origin of Group~IV stars.

We also note that \citet{Hong_et_al_2026} have recently identified a substantial number of additional Group IV candidates using photometric data from the J-PLUS and S-PLUS surveys, indicating that this population is more common than previously recognised.

\subsection{Abundance pattern and nucleosynthetic constraints}

In addition to carbon enhancement, HE~1153$-$0518 exhibits extreme sodium enrichment ([Na/Fe]~$\sim$~+2.1), mild $\alpha$-element enhancement, low manganese, and very low neutron-capture element abundances. Such a combination is difficult to reproduce through classical AGB nucleosynthesis, which is generally accompanied by significant s-process enrichment \citep{Karakas_&_Lattanzio_2014}. 

The [Na/Fe] (= +2.12) abundance ratio is among the highest reported for CEMP-no stars at [Fe/H] $\approx -3.1$ and may therefore provide an important additional constraint on the progenitor mass and explosion energy in faint supernova models \citep{Kobayashi_et_al_2006, Nomoto_et_al_2013}.

Rapidly rotating massive stars (spinstars) have been proposed as possible progenitors of some CEMP-no stars \citep{Meynet_et_al_2006, Meynet_et_al_2010, Frischknecht_et_al_2012, Frischknecht_et_al_2016, maeder_&_meynet_2015, Choplin_et_al_2017}; however, spinstar models typically predict enhanced light $s$-process elements, leading to elevated Sr and Y abundances \citep{Frischknecht_et_al_2016}, which are not observed in HE~1153$-$0518. In contrast, models of faint core-collapse supernovae with mixing and fallback naturally explain the observed pattern of high [C/Fe] and [Na/Fe], low [Fe/H], and low neutron-capture element abundances \citep{Umeda_&_Nomoto_2003, Nomoto_et_al_2013}.

While enrichment by faint core-collapse supernovae remains a viable formation channel for HE~1153$-$0518, its placement in the newly proposed Group IV also allows the possibility that binary mass transfer has contributed to its observed chemical composition, as discussed in Section 6.2.

\subsection{Carbon isotopic ratio, radial-velocity variability, and the role of AGB mass transfer}

The carbon isotopic ratio that we derived for HE~1153$-$0518 is found to be very low ($^{12}$C/$^{13}$C~=~2.0). This value is close to CN-cycle equilibrium and can arise from internal mixing in evolved giants. Similar low isotopic ratios are also observed in some CEMP-$s$ and CEMP-$r/s$ stars and are commonly interpreted as the result of internal mixing operating after the accretion of carbon-rich material from an AGB companion \citep{Spite_et_al_2006, Hansen_et_al_2015}.

From Table~\ref{tab:radial_velocity_table}, we can see that HE~1153$-$0518 might be a radial velocity variable. So, the binary interaction and mass transfer cannot be excluded a priori. Indeed, \citet{Yoon_et_al_2019}, based on the results of \citet{Hansen_et_al_2016} and \citet{Arentsen_et_al_2019}, noted that a significant fraction of high-A(C) CEMP-no stars are binaries, suggesting that AGB mass transfer may operate for at least some members of this population. This possibility is particularly relevant in the context of Group~IV stars, for which \citet{Lee_et_al_2025} have suggested binary mass transfer as a contributing formation channel. However, \citet{Yoon_et_al_2019} also emphasised that binary mass transfer scenario alone cannot simultaneously explain the high carbon abundances and low barium abundances observed in high-A(C) CEMP-no stars. At present, no well-established low-metallicity AGB models can reproduce the observed lack of correlation between A(C) and A(Ba). Thus, while binary interaction with a now extinct AGB may influence the surface composition of HE~1153$-$0518, it may not be the dominant origin of its carbon enrichment. 

\subsection{Evolutionary state, lithium, oxygen, and infrared excess}
\label{ir_excess}

HE~1153$-$0518 has a very low surface gravity (log~g~=~0.20), indicating that it is a highly evolved giant. The non-detection of lithium and the resulting upper limits are consistent with such an evolved evolutionary stage and do not place additional constraints on the formation history \citep{Lind_et_al_2009}. Similarly, the upper limits on oxygen abundance are consistent with those seen in other EMP and CEMP-no stars, and are not decisive enough to distinguish between different early chemical enrichment scenarios \citep{Cayrel_et_al_2004, Spite_et_al_2005}.

The spectral energy distribution of HE~1153$-$0518 shows a clear infrared (IR) excess, indicating the presence of circumstellar dust, which is very common for objects with low surface gravity (log~g~=~0.2). While, such an IR excess might suggest an intrinsic AGB evolutionary phase, the lack of $s$-process enrichment and the very low value of $^{12}$C/$^{13}$C are inconsistent with expectations for thermally pulsing AGB stars.
Recent studies have demonstrated that dust formation can occur even at extremely low metallicities, particularly in carbon-rich environments, without necessarily requiring classical AGB evolution \citep{Boyer_et_al_2015, Ventura_et_al_2021}. In this context, the infrared excess observed in HE~1153$-$0518 provides one of the few direct observational indications supporting dust production at [Fe/H] $< -3.0$.

\subsection{Accretion origin and formation scenario}

Although the Gaia parallax of HE~1153$-$0518 is negative and unreliable, we have estimated its distance photometrically. The derived kinematic parameters, though carrying large uncertainties, place this star firmly in the halo population with P$_{halo}$ = 1.00, which is fully consistent with its extremely low metallicity and CEMP-no chemical abundance pattern.

In this context, HE~1153$-$0518 may have formed in an early and chemically primitive environment. Such environments are commonly found in small systems like ultra-faint dwarf galaxies or dwarf spheroidal galaxies, which were later accreted by the Milky Way \citep{Yoon_et_al_2019}. However, with the present kinematic data of the star, this origin cannot be confirmed. 

The combination of high carbon abundance, strong sodium enhancement, low neutron-capture element abundances, and a carbon isotopic ratio close to CN-cycle equilibrium indicates that enrichment by early massive stars, such as faint core-collapse supernovae, may have played an important role. At the same time, the evidence for binarity and the presence of IR excess suggests that mass transfer from an AGB companion and internal mixing could also have influenced the observed surface abundances. The location of this star in the newly proposed Group~IV further supports the possibility that binary interaction may have contributed to its chemical composition. Further observations, particularly long-term radial-velocity monitoring and improved astrometric data, will be required to better distinguish between these possible formation and enrichment scenarios.

\section{Conclusions}
\label{sec:conclusion}
We have performed detailed high-resolution spectroscopic analysis on three stars: HE~0401$-$0138, HE~1153$-$0518, and HE~1246$-$1344. All three programme stars are found to be extremely metal-poor stars with the [Fe/H] range from --3.08 to --3.50.

\begin{itemize}
    \item The stars HE~0401$-$0138 and HE~1246$-$1344 show chemical abundance patterns typical of normal EMP halo stars, including moderate $\alpha$-element enhancement and low neutron-capture element abundances, consistent with enrichment by early core-collapse supernovae. HE~1153$-$0518 is identified as a newly discovered CEMP-no star belonging to the recently proposed Group~IV. This star shows very high absolute carbon abundance, very low neutron-capture element abundances, and extreme sodium enhancement.

    \item The abundance pattern of HE~1153$-$0518, including extreme sodium enhancement, low neutron-capture elements, and a very low carbon isotopic ratio, suggests that its chemical composition reflects enrichment by material processed through early nucleosynthetic channels. While enrichment by early massive stars, such as faint core-collapse supernovae, provides a natural explanation for several of the observed features, the presence of radial-velocity variability indicates that binary interaction with a now extinct AGB and internal mixing may also have played a role in creating the observed surface abundances. 
    
    \item Although the Gaia parallax of HE~1153$-$0518 is negative and hence not usable directly, we have estimated its distance using a photometric method. The kinematic parameters derived from this estimated distance confirm that HE~1153$-$0518 belongs to the Galactic halo population, consistent with its extremely low metallicity and chemical abundance pattern typical of CEMP-no stars found in the halo. An origin in a small, chemically primitive system such as an ultra-faint dwarf or dwarf spheroidal galaxy, later accreted by the Milky Way, remains a plausible explanation.

\end{itemize}

The discovery and detailed comprehensive study of HE~1153$-$0518 add an important new object to the limited sample of high-A(C) CEMP-no stars. Its unusual combination of strong carbon enhancement, extreme sodium abundance, low neutron-capture element content, low carbon isotopic ratio, and infrared excess makes it a valuable probe for understanding the diversity of early chemical enrichment pathways and the nature of the first generations of stars.

\vspace{0.3cm}
\noindent
{\it \bf{Acknowledgements}}

This work made use of the SIMBAD astronomical database, operated at CDS, Strasbourg, France, the NASA ADS, USA and data from the European Space Agency (ESA) mission Gaia (\url{https://www.cosmos.esa.int/gaia}), processed by the Gaia Data Processing and Analysis Consortium (DPAC, \url{https://www.cosmos.esa.int/web/gaia/dpac/consortium}). This publication makes use of VOSA, developed under the Spanish Virtual Observatory (\url{https://svo.cab.inta-csic.es}) project funded by MCIN/AEI/10.13039/501100011033/ through grant PID2020-112949GB-I00. VOSA has been partially updated by using funding from the European Union's Horizon 2020 Research and Innovation Programme, under Grant Agreement nº 776403 (EXOPLANETS-A).

\section{Data availability}

The spectroscopic data underlying this article were obtained from the Subaru Telescope archive through the JVO portal (http://jvo.nao.ac.jp/portal) operated by the National Astronomical Observatory of Japan (NAOJ). Additional photometric and astrometric data used in this work were obtained from publicly available databases including SIMBAD and Gaia. The derived abundance data and other results are included within the article. Additional data products will be shared on reasonable request to the corresponding author.

\bibliographystyle{mnras}
\bibliography{sample}

\appendix

{\footnotesize
\begin{table}
\centering
\caption{\bf{Equivalent widths (in m\r{A}) of lines used for calculation of elemental abundances.}}
\label{tab:Elem_linelist1_appendix} 
\scalebox{0.75}{
\begin{tabular}{cccccccc}
\hline
Wavelength   &Element    &E$_{low}$ &   log gf  &  HE~0401$-$0138  &  HE~1246$-$1344  & HE~1153$-$0518   \\
(\r{A})      &           & (eV)     &           &                 &                 &                 \\
\hline 
4045.81      &  Fe I     &  1.48    &    0.280  &   101.3 (4.10)  &      -          &      -          \\
4198.30      &           &  2.40    & $-$0.719  &        -        &   46.2 (4.16)   &      -          \\
4202.03      &           &  1.48    & $-$0.708  &        -        &   78.6 (3.85)   &      -          \\
4233.60      &           &  2.48    & $-$0.604  &        -        &   33.6 (3.87)   &      -          \\
4250.79      &           &  1.56    & $-$0.710  &    64.9 (3.82)  &      -          &      -          \\
4271.76      &           &  1.48    & $-$0.164  &        -        &  102.1 (4.01)   &      -          \\
4375.93      &           &  0.00    & $-$3.031  &    59.9 (3.98)  &      -          &      -          \\
4383.55      &           &  1.48    &    0.200  &    96.9 (3.86)  &  114.5 (3.94)   &      -          \\
4404.75      &           &  1.56    & $-$0.142  &    88.9 (4.02)  &   97.8 (3.86)   &      -          \\
4415.12      &           &  1.61    & $-$0.615  &    73.7 (4.03)  &   87.8 (4.10)   &      -          \\
4427.31      &           &  0.05    & $-$3.044  &    68.8 (4.31)  &   62.2 (3.84)   &      -          \\
4442.34      &           &  2.20    & $-$1.255  &    21.5 (3.93)  &   35.2 (4.18)   &      -          \\
4447.72      &           &  2.22    & $-$1.342  &    24.8 (4.13)  &      -          &      -          \\
4466.573     &           &  0.11    & $-$4.464  &       -         &      -          &      -          \\
4528.61      &           &  2.18    & $-$0.822  &       -         &   41.0 (3.83)   &      -          \\ 
4602.94      &           &  1.48    & $-$1.950  &       -         &      -          &      -          \\
4871.32      &           &  2.87    & $-$0.410  &       -         &   31.1 (4.00)   &      -          \\
4872.14      &           &  2.88    & $-$0.600  &       -         &   22.3 (3.99)   &      -          \\
4890.76      &           &  2.88    & $-$0.430  &       -         &   31.4 (4.03)   &      -          \\
4891.49      &           &  2.85    & $-$0.140  &    42.5 (4.05)  &   48.5 (4.05)   &      -          \\
4918.99      &           &  2.87    & $-$0.370  &    34.0 (4.09)  &   28.5 (3.90)   &      -          \\
4920.50      &           &  2.83    &    0.060  &    46.7 (3.92)  &   49.7 (3.85)   &    71.2 (4.12)  \\
4982.50      &           &  4.10    &    0.164  &       -         &      -          &      -          \\
4994.13      &           &  0.92    & $-$3.080  &       -         &      -          &      -          \\
5006.12      &           &  2.83    & $-$0.615  &       -         &   22.6 (3.94)   &      -          \\
5041.76      &           &  1.48    & $-$2.203  &       -         &   32.2 (4.11)   &      -          \\
5049.82      &           &  2.28    & $-$1.420  &       -         &   19.9 (4.02)   &      -          \\
5051.64      &           &  0.91    & $-$2.795  &    30.3 (4.05)  &   31.6 (3.98)   &    85.3 (4.74)  \\
5083.34      &           &  0.96    & $-$2.958  &    21.4 (4.05)  &   20.9 (3.96)   &      -          \\
5166.28      &           &  0.00    & $-$4.195  &    16.6 (3.93)  &      -          &      -          \\
5171.60      &           &  1.49    & $-$1.793  &    44.2 (4.06)  &   44.6 (3.94)   &      -          \\
5192.34      &           &  3.00    & $-$0.421  &       -         &   21.8 (3.91)   &    57.6 (4.51)  \\
5194.94      &           &  1.56    & $-$2.090  &       -         &   37.7 (4.19)   &    71.0 (4.60)  \\
5202.34      &           &  2.18    & $-$1.838  &    11.6 (4.07)  &      -          &      -          \\
5216.27      &           &  1.61    & $-$2.150  &       -         &   28.0 (4.11)   &    56.4 (4.47)  \\
5217.39      &           &  3.21    & $-$1.162  &       -         &      -          &      -          \\
5226.86      &           &  3.04    & $-$0.555  &    17.5 (4.02)  &      -          &      -          \\
5227.19      &           &  1.56    & $-$1.228  &    72.8 (4.25)  &   72.3 (3.98)   &   108.5 (4.47)  \\
5232.94      &           &  2.94    & $-$0.190  &    35.5 (4.00)  &   33.6 (3.88)   &    70.6 (4.44)  \\
5242.49      &           &  3.63    & $-$0.840  &       -         &      -          &      -          \\
5266.56      &           &  3.00    & $-$0.490  &    21.4 (4.03)  &   21.2 (3.96)   &    39.1 (4.25)  \\
5269.54      &           &  0.86    & $-$1.321  &    85.7 (3.81)  &  105.8 (3.95)   &   133.7 (4.19)  \\
5270.36      &           &  1.61    & $-$1.510  &    62.1 (4.32)  &   64.1 (4.15)   &    95.8 (4.53)  \\
5281.79      &           &  3.04    & $-$1.020  &    14.1 (4.37)  &      -          &      -          \\
5283.62      &           &  3.24    & $-$0.630  &       -         &      -          &    38.8 (4.67)  \\
5324.18      &           &  3.21    & $-$0.240  &    20.2 (3.98)  &   20.9 (3.95)   &      -          \\
5328.04      &           &  0.91    & $-$1.466  &    83.2 (3.93)  &  103.3 (4.06)   &   123.2 (4.11)  \\
5328.53      &           &  1.56    & $-$1.850  &    36.6 (4.02)  &   44.0 (4.05)   &    75.3 (4.40)  \\
5341.02      &           &  1.61    & $-$2.060  &       -         &   28.5 (4.02)   &      -          \\
5446.92      &           &  0.99    & $-$1.930  &       -         &   77.6 (4.01)   &      -          \\
5455.61      &           &  1.01    & $-$2.091  &    64.8 (4.17)  &   67.4 (4.00)   &   119.3 (4.74)  \\
5497.52      &           &  1.01    & $-$2.849  &    24.5 (4.04)  &   27.9 (4.03)   &      -          \\
5501.47      &           &  0.96    & $-$2.950  &       -         &   18.7 (3.84)   &      -          \\
5506.78      &           &  0.99    & $-$2.797  &    32.8 (4.16)  &   29.8 (3.99)   &    44.7 (4.09)  \\
5572.84      &           &  3.40    & $-$0.310  &    15.5 (4.11)  &   13.8 (4.00)   &      -          \\
5586.76      &           &  3.37    & $-$0.210  &       -         &   18.3 (4.01)   &      -          \\
5615.64      &           &  3.33    & $-$0.140  &       -         &   24.6 (4.06)   &      -          \\
6136.61      &           &  2.45    & $-$1.400  &     9.7 (3.80)  &      -          &      -          \\
6230.72      &           &  2.56    & $-$1.281  &       -         &      -          &      -          \\
6335.33      &           &  2.20    & $-$2.230  &       -         &      -          &    15.8 (4.38)  \\
6494.98      &           &  2.40    & $-$1.273  &    21.0 (3.99)  &   21.6 (3.94)   &      -          \\
4233.17      &  Fe II    &  2.58    & $-$2.000  &    21.0 (4.01)  &      -          &      -          \\
4555.89      &           &  2.83    & $-$2.290  &       -         &   10.5 (4.01)   &      -          \\
4583.84      &           &  2.81    & $-$2.020  &       -         &      -          &      -          \\
4923.93      &           &  2.89    & $-$1.320  &    37.3 (4.14)  &      -          &    67.7 (4.30)  \\
5018.44      &           &  2.89    & $-$1.220  &    36.9 (4.04)  &      -          &      -          \\
5197.58      &           &  3.23    & $-$2.100  &     6.1 (4.10)  &      -          &    27.9 (4.53)  \\
5234.62      &           &  3.22    & $-$2.050  &       -         &    7.7 (4.02)   &      -          \\
5276.00      &           &  3.20    & $-$1.940  &       -         &      -          &      -          \\
5889.95      &  Na I     &  0.00    &    0.117  &    93.1 (3.20)  &  146.6 (4.26)   &   244.5 (5.20)  \\
5895.92      &           &  0.00    & $-$0.184  &    71.5 (2.94)  &  133.0 (4.28)   &   225.2 (5.36)  \\
4571.10      &  Mg I     &  0.00    & $-$5.691  &       -         &   16.7 (4.68)   &      -          \\
5172.68      &           &  2.71    & $-$0.402  &   114.6 (5.01)  &      -          &   156.5 (5.58)  \\
5183.60      &           &  2.72    & $-$0.180  &   132.1 (5.14)  &      -          &   169.5 (5.55)  \\
5528.40      &           &  4.35    & $-$0.620  &    26.6 (4.89)  &   31.0 (4.97)   &      -          \\
4226.73      &  Ca I     &  0.00    &    0.243  &       -         &  108.3 (3.42)   &      -          \\
4283.01      &           &  1.89    & $-$0.224  &       -         &      -          &      -          \\
4302.53      &           &  1.90    &    0.275  &       -         &   32.7 (3.29)   &      -          \\
4318.65      &           &  1.90    & $-$0.208  &    19.4 (3.41)  &   16.7 (3.31)   &      -          \\
4434.96      &           &  1.89    & $-$0.029  &       -         &   28.2 (3.44)   &      -          \\
4454.78      &           &  1.90    &    0.252  &       -         &   36.9 (3.38)   &      -          \\
5588.75      &           &  2.53    &    0.210  &       -         &   17.8 (3.52)   &      -          \\
5594.46      &           &  2.52    & $-$0.050  &       -         &      -          &      -          \\
5857.45      &           &  2.93    &    0.230  &       -         &      -          &   19.4 (3.92)   \\
6102.72      &           &  1.88    & $-$0.890  &       -         &      -          &      -          \\
6122.22      &           &  1.89    & $-$0.409  &    19.0 (3.43)  &      -          &      -          \\
\hline
\end{tabular}}

\end{table}}

{\footnotesize
\begin{table}
\centering
{\bf{--\it{continued}}}
\scalebox{0.75}{
\begin{tabular}{ccccccc}
\hline
Wavelength   &Element    &E$_{low}$ &   log gf  &  HE~0401$-$0138  &  HE~1246$-$1344  & HE~1153$-$0518   \\
(\r{A})      &           & (eV)     &           &                 &                 &                 \\
\hline 
6162.17      &           &  1.90    &    0.100  &       -         &      -          &      -          \\
6439.07      &           &  2.53    &    0.470  &    22.7 (3.38)  &   14.0 (3.07)   &      -          \\
4533.24      &  Ti I     &  0.85    &    0.476  &    14.0 (1.84)  &      -          &      -          \\
4981.73      &           &  0.85    &    0.504  &    13.7 (1.75)  &   16.0 (1.81)   &      -          \\
4991.06      &           &  0.84    &    0.380  &       -         &      -          &      -          \\
4999.50      &           &  0.83    &    0.250  &       -         &   11.9 (1.88)   &      -          \\
5007.21      &           &  0.82    &    0.112  &       -         &      -          &      -          \\
5192.97      &           &  0.02    & $-$1.006  &    7.1 (1.93)   &      -          &   28.5 (2.56)   \\
4028.34      &  Ti II    &  1.89    & $-$1.000  &       -         &      -          &      -          \\
4290.22      &           &  1.16    & $-$1.120  &    52.4 (2.24)  &   45.6 (1.79)   &      -          \\
4300.05      &           &  1.18    & $-$0.770  &       -         &      -          &      -          \\
4312.86      &           &  1.18    & $-$1.160  &       -         &      -          &      -          \\
4395.03      &           &  1.08    & $-$0.660  &       -         &  72.2 (2.03)    &      -          \\
4417.72      &           &  1.16    & $-$1.430  &       -         &  40.1 (1.92)    &      -          \\
4443.79      &           &  1.08    & $-$0.700  &    63.5 (2.09)  &  69.1 (1.94)    &   90.1 (2.23)   \\
4444.56      &           &  1.12    & $-$2.030  &     8.8 (1.63)  &      -          &      -          \\
4450.48      &           &  1.08    & $-$1.450  &       -         &  35.4 (1.73)    &      -          \\
4468.51      &           &  1.13    & $-$0.600  &       -         &  64.1 (1.72)    &      -          \\
4470.86      &           &  1.16    & $-$2.280  &     8.2 (1.88)  &      -          &      -          \\
4501.27      &           &  1.12    & $-$0.750  &    63.6 (2.17)  &  63.6 (1.83)    &      -          \\
4533.97      &           &  1.24    & $-$0.770  &       -         &  61.6 (1.93)    &      -          \\
4563.76      &           &  1.22    & $-$0.960  &    47.6 (1.91)  &  58.6 (1.99)    &      -          \\
4571.97      &           &  1.57    & $-$0.530  &       -         &      -          &      -          \\
4589.96      &           &  1.24    & $-$1.790  &       -         &  22.6 (1.91)    &      -          \\
4779.98      &           &  2.05    & $-$1.370  &       -         &      -          &   38.7 (2.54)   \\
4805.09      &           &  2.06    & $-$1.100  &       -         &      -          &      -          \\
4865.61      &           &  1.12    & $-$2.610  &       -         &      -          &   19.5 (2.18)   \\
5129.15      &           &  1.89    & $-$1.390  &       -         &      -          &      -          \\
5185.91      &           &  1.89    & $-$1.350  &       -         &      -          &      -          \\
5188.68      &           &  1.58    & $-$1.210  &       -         &  29.1 (1.82)    &      -          \\
5226.54      &           &  1.57    & $-$1.300  &       -         &  23.6 (1.76)    &   65.1 (2.38)   \\
4254.34      &  Cr I     &  0.00    & $-$0.114  &    55.4 (2.08)  &  55.7 (1.88)    &      -          \\
4274.80      &           &  0.00    & $-$0.231  &       -         &  45.2 (1.65)    &      -          \\
4289.72      &           &  0.00    & $-$0.361  &    52.5 (2.20)  &  54.3 (2.06)    &      -          \\
5206.04      &           &  0.94    &    0.019  &       -         &  27.1 (1.88)    &   65.0 (2.70)   \\
4092.38      &  Co I     &  0.92    & $-$0.940  &       -         &      -          &      -          \\
4118.77      &           &  1.05    & $-$0.490  &       -         &  14.5 (1.62)    &      -          \\
4121.31      &           &  0.92    & $-$0.320  &    27.7 (1.76)  &      -          &      -          \\
4714.41      &  Ni I     &  3.38    &    0.260  &       -         &   5.7 (2.86)    &      -          \\
5476.90      &           &  1.83    & $-$0.890  &    22.0 (2.89)  &  18.6 (2.74)    &   48.4 (3.35)   \\
4077.71      &  Sr II    &  0.00    &    0.167  &    78.2 (-0.39) &  57.0 (-1.62)   &      -          \\
4215.52      &           &  0.00    & $-$0.145  &       -         &  46.3 (-1.78)   &      -          \\
4177.53      &  Y II     &  0.41    & $-$0.160  &    25.9 (-0.90) &     -           &      -          \\
4900.12      &           &  1.03    & $-$0.090  &       -         &     -           &   23.3 (--0.89) \\
\hline
\end{tabular}}

The numbers in the parenthesis in columns 5 to 7 give the derived abundances from the respective line.\\. 

\end{table}}

\label{lastpage}

\end{document}